\documentclass[a4paper,11pt]{article}
\usepackage{jcappub}
\usepackage{hyperref}
\usepackage{verbatim}
\usepackage{tabularx}
\usepackage{breqn}
\usepackage{aas_macros}
\usepackage{orcidlink}

\newcommand{\otwo}{{[O\,II]\,}}

\def\D{{\Delta}}
\def\L{{\Lambda}}

\def\2pr{^{\prime \prime}}

\def\geqsim{\lower.73ex\hbox{$\sim$}\llap{\raise.4ex\hbox{$>$}}$\,$}
\def\leqsim{\lower.73ex\hbox{$\sim$}\llap{\raise.4ex\hbox{$<$}}$\,$}

\title{The Construction of Large-scale Structure Catalogs for the Dark Energy Spectroscopic Instrument  }

\author[1]{{A.~J.~Ross}\orcidlink{0000-0002-7522-9083},}
\author[2]{{J.~Aguilar},}
\author[3]{{S.~Ahlen}\orcidlink{0000-0001-6098-7247},}
\author[4]{{S.~Alam}\orcidlink{0000-0002-3757-6359},}
\author[2]{{A.~Anand}\orcidlink{0000-0003-2923-1585},}
\author[2]{{S.~Bailey}\orcidlink{0000-0003-4162-6619},}
\author[5]{{D.~Bianchi}\orcidlink{0000-0001-9712-0006},}
\author[6]{{S.~Brieden}\orcidlink{0000-0003-3896-9215},}
\author[7]{{D.~Brooks},}
\author[8]{{E.~Burtin},}
\author[9,10]{{A.~Carnero Rosell}\orcidlink{0000-0003-3044-5150},}
\author[2]{{E.~Chaussidon}\orcidlink{0000-0001-8996-4874},}
\author[2]{{T.~Claybaugh},}
\author[11]{{S.~Cole}\orcidlink{0000-0002-5954-7903},}
\author[12]{{K.~Dawson},}
\author[13]{{A.~de la Macorra}\orcidlink{0000-0002-1769-1640},}
\author[8]{{A.~de~Mattia}\orcidlink{0000-0003-0920-2947},}
\author[14]{{Arjun~Dey}\orcidlink{0000-0002-4928-4003},}
\author[15]{{Biprateep~Dey}\orcidlink{0000-0002-5665-7912},}
\author[7]{{P.~Doel},}
\author[16,17]{{K.~Fanning}\orcidlink{0000-0003-2371-3356},}
\author[2,18]{{S.~Ferraro}\orcidlink{0000-0003-4992-7854},}
\author[19]{{J.~Ereza}\orcidlink{0000-0002-0194-4017},}
\author[7,20]{{A.~Font-Ribera}\orcidlink{0000-0002-3033-7312},}
\author[21,22]{{J.~E.~Forero-Romero}\orcidlink{0000-0002-2890-3725},}
\author[23,24,25]{{E.~Gaztañaga},}
\author[26,23,27]{{H.~Gil-Mar\'in}\orcidlink{0000-0003-0265-6217},}
\author[2]{{S.~Gontcho A Gontcho}\orcidlink{0000-0003-3142-233X},}
\author[28,29]{{A.~X.~Gonzalez-Morales}\orcidlink{0000-0003-4089-6924},}
\author[2]{{J.~Guy}\orcidlink{0000-0001-9822-6793},}
\author[30]{{C.~Hahn}\orcidlink{0000-0003-1197-0902},}
\author[31]{{S.~Heydenreich},}
\author[1,32,33]{{K.~Honscheid},}
\author[34]{{C.~Howlett}\orcidlink{0000-0002-1081-9410},}
\author[35]{{M.~Ishak}\orcidlink{0000-0002-6024-466X},}
\author[36,37]{{T.~Karim}\orcidlink{0000-0002-5652-8870},}
\author[38]{{D.~Kirkby}\orcidlink{0000-0002-8828-5463},}
\author[2]{{T.~Kisner}\orcidlink{0000-0003-3510-7134},}
\author[20,33]{{H.~Kong},}
\author[2]{{A.~Kremin}\orcidlink{0000-0001-6356-7424},}
\author[39,40,41]{{A.~Krolewski},}
\author[2]{{A.~Lambert},}
\author[2]{{M.~Landriau}\orcidlink{0000-0003-1838-8528},}
\author[42]{{J.~Lasker}\orcidlink{0000-0003-2999-4873},}
\author[43]{{L.~Le~Guillou}\orcidlink{0000-0001-7178-8868},}
\author[2]{{M.~E.~Levi}\orcidlink{0000-0003-1887-1018},}
\author[44,20]{{M.~Manera}\orcidlink{0000-0003-4962-8934},}
\author[1,45,33]{{P.~Martini}\orcidlink{0000-0002-4279-4182},}
\author[2]{{P.~McDonald}\orcidlink{0000-0001-8346-8394},}
\author[14]{{A.~Meisner}\orcidlink{0000-0002-1125-7384},}
\author[46,20]{{R.~Miquel},}
\author[47]{{J.~Moon},}
\author[48]{{J.~Moustakas}\orcidlink{0000-0002-2733-4559},}
\author[13]{{A.~Muñoz-Gutiérrez},}
\author[49]{{A.~D.~Myers},}
\author[24]{{S.~Nadathur}\orcidlink{0000-0001-9070-3102},}
\author[49]{{L.~Napolitano}\orcidlink{0000-0002-5166-8671},}
\author[15]{{J.~ A.~Newman}\orcidlink{0000-0001-8684-2222},}
\author[50]{{J.~Nie}\orcidlink{0000-0001-6590-8122},}
\author[29,51]{{G.~Niz}\orcidlink{0000-0002-1544-8946},}
\author[8,2]{{N.~Palanque-Delabrouille}\orcidlink{0000-0003-3188-784X},}
\author[39,40,41]{{W.~J.~Percival}\orcidlink{0000-0002-0644-5727},}
\author[2,52,18]{{C.~Poppett},}
\author[19]{{F.~Prada}\orcidlink{0000-0001-7145-8674},}
\author[2]{{A.~Raichoor}\orcidlink{0000-0001-5999-7923},}
\author[53,8,54]{{C.~Ravoux}\orcidlink{0000-0002-3500-6635},}
\author[55]{{M.~Rezaie}\orcidlink{0000-0001-5589-7116},}
\author[56]{{A.~Rosado-Marin},}
\author[57]{{G.~Rossi},}
\author[58,55,59]{{L.~Samushia}\orcidlink{0000-0002-1609-5687},}
\author[60]{{E.~Sanchez}\orcidlink{0000-0002-9646-8198},}
\author[61]{{E.~F.~Schlafly}\orcidlink{0000-0002-3569-7421},}
\author[2]{{D.~Schlegel},}
\author[56]{{H.~Seo}\orcidlink{0000-0002-6588-3508},}
\author[11]{{A.~Smith}\orcidlink{0000-0002-3712-6892},}
\author[14]{{D.~Sprayberry},}
\author[62]{{G.~Tarl\'{e}}\orcidlink{0000-0003-1704-0781},}
\author[56]{{D.~Valcin}\orcidlink{0000-0003-0129-0620},}
\author[13]{{M.~Vargas-Maga\~na}\orcidlink{0000-0003-3841-1836},}
\author[14]{{B.~A.~Weaver},}
\author[11]{{Michael~J.~Wilson},}
\author[63]{{J.~Yu}\orcidlink{0009-0001-7217-8006},}
\author[43]{{P.~Zarrouk}\orcidlink{0000-0002-7305-9578},}
\author[64]{{C.~Zhao}\orcidlink{0000-0002-1991-7295},}
\author[2]{{R.~Zhou}\orcidlink{0000-0001-5381-4372},}
\author[50]{{H.~Zou}\orcidlink{0000-0002-6684-3997},}

\affiliation[1]{Center for Cosmology and AstroParticle Physics, The Ohio State University, 191 West Woodruff Avenue, Columbus, OH 43210, USA}
\affiliation[2]{Lawrence Berkeley National Laboratory, 1 Cyclotron Road, Berkeley, CA 94720, USA}
\affiliation[3]{Physics Dept., Boston University, 590 Commonwealth Avenue, Boston, MA 02215, USA}
\affiliation[4]{Tata Institute of Fundamental Research, Homi Bhabha Road, Mumbai 400005, India}
\affiliation[5]{Dipartimento di Fisica ``Aldo Pontremoli'', Universit\`a degli Studi di Milano, Via Celoria 16, I-20133 Milano, Italy}
\affiliation[6]{Institute for Astronomy, University of Edinburgh, Royal Observatory, Blackford Hill, Edinburgh EH9 3HJ, UK}
\affiliation[7]{Department of Physics \& Astronomy, University College London, Gower Street, London, WC1E 6BT, UK}
\affiliation[8]{IRFU, CEA, Universit\'{e} Paris-Saclay, F-91191 Gif-sur-Yvette, France}
\affiliation[9]{Departamento de Astrof\'{\i}sica, Universidad de La Laguna (ULL), E-38206, La Laguna, Tenerife, Spain}
\affiliation[10]{Instituto de Astrof\'{\i}sica de Canarias, C/ V\'{\i}a L\'{a}ctea, s/n, E-38205 La Laguna, Tenerife, Spain}
\affiliation[11]{Institute for Computational Cosmology, Department of Physics, Durham University, South Road, Durham DH1 3LE, UK}
\affiliation[12]{Department of Physics and Astronomy, The University of Utah, 115 South 1400 East, Salt Lake City, UT 84112, USA}
\affiliation[13]{Instituto de F\'{\i}sica, Universidad Nacional Aut\'{o}noma de M\'{e}xico,  Cd. de M\'{e}xico  C.P. 04510,  M\'{e}xico}
\affiliation[14]{NSF NOIRLab, 950 N. Cherry Ave., Tucson, AZ 85719, USA}
\affiliation[15]{Department of Physics \& Astronomy and Pittsburgh Particle Physics, Astrophysics, and Cosmology Center (PITT PACC), University of Pittsburgh, 3941 O'Hara Street, Pittsburgh, PA 15260, USA}
\affiliation[16]{Kavli Institute for Particle Astrophysics and Cosmology, Stanford University, Menlo Park, CA 94305, USA}
\affiliation[17]{SLAC National Accelerator Laboratory, Menlo Park, CA 94305, USA}
\affiliation[18]{University of California, Berkeley, 110 Sproul Hall \#5800 Berkeley, CA 94720, USA}
\affiliation[19]{Instituto de Astrof\'{i}sica de Andaluc\'{i}a (CSIC), Glorieta de la Astronom\'{i}a, s/n, E-18008 Granada, Spain}
\affiliation[20]{Institut de F\'{i}sica d’Altes Energies (IFAE), The Barcelona Institute of Science and Technology, Campus UAB, 08193 Bellaterra Barcelona, Spain}
\affiliation[21]{Departamento de F\'isica, Universidad de los Andes, Cra. 1 No. 18A-10, Edificio Ip, CP 111711, Bogot\'a, Colombia}
\affiliation[22]{Observatorio Astron\'omico, Universidad de los Andes, Cra. 1 No. 18A-10, Edificio H, CP 111711 Bogot\'a, Colombia}
\affiliation[23]{Institut d'Estudis Espacials de Catalunya (IEEC), 08034 Barcelona, Spain}
\affiliation[24]{Institute of Cosmology and Gravitation, University of Portsmouth, Dennis Sciama Building, Portsmouth, PO1 3FX, UK}
\affiliation[25]{Institute of Space Sciences, ICE-CSIC, Campus UAB, Carrer de Can Magrans s/n, 08913 Bellaterra, Barcelona, Spain}
\affiliation[26]{Departament de F\'{\i}sica Qu\`{a}ntica i Astrof\'{\i}sica, Universitat de Barcelona, Mart\'{\i} i Franqu\`{e}s 1, E08028 Barcelona, Spain}
\affiliation[27]{Institut de Ci\`encies del Cosmos (ICCUB), Universitat de Barcelona (UB), c. Mart\'i i Franqu\`es, 1, 08028 Barcelona, Spain.}
\affiliation[28]{Consejo Nacional de Ciencia y Tecnolog\'{\i}a, Av. Insurgentes Sur 1582. Colonia Cr\'{e}dito Constructor, Del. Benito Ju\'{a}rez C.P. 03940, M\'{e}xico D.F. M\'{e}xico}
\affiliation[29]{Departamento de F\'{i}sica, Universidad de Guanajuato - DCI, C.P. 37150, Leon, Guanajuato, M\'{e}xico}
\affiliation[30]{Department of Astrophysical Sciences, Princeton University, Princeton NJ 08544, USA}
\affiliation[31]{Department of Astronomy and Astrophysics, UCO/Lick Observatory, University of California, 1156 High Street, Santa Cruz, CA 95064, USA}
\affiliation[32]{Department of Physics, The Ohio State University, 191 West Woodruff Avenue, Columbus, OH 43210, USA}
\affiliation[33]{The Ohio State University, Columbus, 43210 OH, USA}
\affiliation[34]{School of Mathematics and Physics, University of Queensland, 4072, Australia}
\affiliation[35]{Department of Physics, The University of Texas at Dallas, Richardson, TX 75080, USA}
\affiliation[36]{Center for Astrophysics $|$ Harvard \& Smithsonian, 60 Garden Street, Cambridge, MA 02138, USA}
\affiliation[37]{Department of Astronomy \& Astrophysics, University of Toronto, Toronto, ON M5S 3H4, Canada}
\affiliation[38]{Department of Physics and Astronomy, University of California, Irvine, 92697, USA}
\affiliation[39]{Department of Physics and Astronomy, University of Waterloo, 200 University Ave W, Waterloo, ON N2L 3G1, Canada}
\affiliation[40]{Perimeter Institute for Theoretical Physics, 31 Caroline St. North, Waterloo, ON N2L 2Y5, Canada}
\affiliation[41]{Waterloo Centre for Astrophysics, University of Waterloo, 200 University Ave W, Waterloo, ON N2L 3G1, Canada}
\affiliation[42]{Department of Physics, Southern Methodist University, 3215 Daniel Avenue, Dallas, TX 75275, USA}
\affiliation[43]{Sorbonne Universit\'{e}, CNRS/IN2P3, Laboratoire de Physique Nucl\'{e}aire et de Hautes Energies (LPNHE), FR-75005 Paris, France}
\affiliation[44]{Departament de F\'{i}sica, Serra H\'{u}nter, Universitat Aut\`{o}noma de Barcelona, 08193 Bellaterra (Barcelona), Spain}
\affiliation[45]{Department of Astronomy, The Ohio State University, 4055 McPherson Laboratory, 140 W 18th Avenue, Columbus, OH 43210, USA}
\affiliation[46]{Instituci\'{o} Catalana de Recerca i Estudis Avan\c{c}ats, Passeig de Llu\'{\i}s Companys, 23, 08010 Barcelona, Spain}
\affiliation[47]{Max Planck Institute for Extraterrestrial Physics, Gie\ss enbachstra\ss e 1, 85748 Garching, Germany}
\affiliation[48]{Department of Physics and Astronomy, Siena College, 515 Loudon Road, Loudonville, NY 12211, USA}
\affiliation[49]{Department of Physics \& Astronomy, University  of Wyoming, 1000 E. University, Dept.~3905, Laramie, WY 82071, USA}
\affiliation[50]{National Astronomical Observatories, Chinese Academy of Sciences, A20 Datun Rd., Chaoyang District, Beijing, 100012, P.R. China}
\affiliation[51]{Instituto Avanzado de Cosmolog\'{\i}a A.~C., San Marcos 11 - Atenas 202. Magdalena Contreras, 10720. Ciudad de M\'{e}xico, M\'{e}xico}
\affiliation[52]{Space Sciences Laboratory, University of California, Berkeley, 7 Gauss Way, Berkeley, CA  94720, USA}
\affiliation[53]{Aix Marseille Univ, CNRS/IN2P3, CPPM, Marseille, France}
\affiliation[54]{Universit\'{e} Clermont-Auvergne, CNRS, LPCA, 63000 Clermont-Ferrand, France}
\affiliation[55]{Department of Physics, Kansas State University, 116 Cardwell Hall, Manhattan, KS 66506, USA}
\affiliation[56]{Department of Physics \& Astronomy, Ohio University, Athens, OH 45701, USA}
\affiliation[57]{Department of Physics and Astronomy, Sejong University, Seoul, 143-747, Korea}
\affiliation[58]{Abastumani Astrophysical Observatory, Tbilisi, GE-0179, Georgia}
\affiliation[59]{Faculty of Natural Sciences and Medicine, Ilia State University, 0194 Tbilisi, Georgia}
\affiliation[60]{CIEMAT, Avenida Complutense 40, E-28040 Madrid, Spain}
\affiliation[61]{Space Telescope Science Institute, 3700 San Martin Drive, Baltimore, MD 21218, USA}
\affiliation[62]{University of Michigan, Ann Arbor, MI 48109, USA}
\affiliation[63]{Ecole Polytechnique F\'{e}d\'{e}rale de Lausanne, CH-1015 Lausanne, Switzerland}
\affiliation[64]{Department of Astronomy, Tsinghua University, 30 Shuangqing Road, Haidian District, Beijing, China, 100190}

\abstract{We present the technical details on how large-scale structure (LSS) catalogs are constructed from redshifts measured from spectra observed by the Dark Energy Spectroscopic Instrument (DESI). The LSS catalogs provide the information needed to determine the relative number density of DESI tracers as a function of redshift and celestial coordinates and, e.g., determine clustering statistics. We produce catalogs that are weighted subsamples of the observed data, each matched to a weighted `random' catalog that forms an unclustered sampling of the probability density that DESI could have observed those data at each location.
 Precise knowledge of the DESI observing history and associated hardware performance allows for a determination of the DESI footprint and the number of times DESI has covered it at sub-arcsecond level precision. This enables the completeness of any DESI sample to be modeled at this same resolution.
The pipeline developed to create LSS catalogs has been designed to easily allow robustness tests and enable future improvements. We describe how it allows ongoing work improving the match between galaxy and random catalogs, such as including further information when assigning redshifts to randoms, accounting for fluctuations in target density, accounting for variation in the redshift success rate, and accommodating blinding schemes.}
\begin{document}

\maketitle

\flushbottom

\section{Introduction}

The Dark Energy Spectroscopic Instrument (DESI; \cite{DESIoverview}) was designed to construct three-dimensional maps of the structure of the Universe. It does so by measuring the spectra of pre-selected galaxy and quasar `targets', from which redshifts can be measured. Given a cosmological model, the angular celestial coordinates and redshifts can be transformed to comoving Cartesian coordinates. The resulting distribution (or, more commonly, its summary statistics) can be compared to that predicted by the cosmological model. In this manner, cosmological constraints obtained with DESI, and the results from Data Release 1 (DR1;\cite{DR1}) are presented in \cite{DESI2024.III.KP4,DESI2024.IV.KP6,DESI2024.V.KP5,DESI2024.VI.KP7A,DESI2024.VII.KP7B,DESI2024.VIII.KP7C}.

Properly interpreting the structure traced by the celestial coordinates and redshifts requires additional context: The relative likelihood as a function of target properties, celestial coordinates, and redshift that our instrument could have successfully measured a redshift. This is commonly referred to as the `selection function' and it is typically not isotropic and has a complicated redshift dependence. To account for this, we produce `Large-Scale structure (LSS) catalogs', which pair the observed data with `randoms' that Poisson sample the observed space. 
The match between the galaxy and random catalogs is achieved by applying weights to the data and randoms and also sub-sampling the randoms, thereby accounting for effects such as the survey footprint, completeness variations due to the number of times DESI has covered the footprint, instrument performance, and systematic variation due to observing conditions.

DESI builds on the tradition of the Sloan Digital Sky Survey (SDSS) LSS catalogs, which are summarized in \cite{ebosslss}. Early versions of the DESI LSS catalog pipeline were applied to `Survey Validation' (SV) data \cite{DESI2023a.KP1.SV} from the DESI early data release (EDR \cite{EDR}) and an early processing of the data from the first two months of the main survey observing \cite{Moon23}. In this work, we provide a technical description of the DESI LSS pipeline, as applied to DESI data release one (DR1; \cite{DR1}). The process and driving philosophy are meant to be constant throughout DESI data releases. However, specifics are likely to change as methods improve. The data model is expected to change slightly to reflect this and will be updated with each data release\footnote{The data model is here: \url{https://desidatamodel.readthedocs.io/en/latest/DESI_ROOT/vac/RELEASE/lss/index.html}}. We focus on the structure of the pipeline and discuss how it is built with flexibility in mind so that improved algorithms can be used in the future.
In many cases, we will reference the specific choices adopted in DR1, but this is meant only to provide concrete examples. The motivation and validation of the specific choices for DR1 are given in \cite{DESI2024.II.KP3} and its companion papers. 

The aspect that is least likely to change is how we track the angular celestial coordinates where DESI observations were possible. We thus begin this work in Section \ref{sec:instrument} by discussing the DESI instrument, with a focus on the focal plane geometry, as mapped onto celestial coordinates, and the strategy with which observations are combined to cover the sky. We then detail the steps of the LSS catalog pipeline: 

\noindent$\bullet$ In Section \ref{sec:inputs}, we describe the initial steps to gather all of the information on potentially assigned targets, observed spectra, and randoms associated with each DESI observation. We describe how it is combined, with each target appearing potentially many times, once for every combination of telescope pointing and hardware configuration that it was observable. We then describe how this information is used to track the DESI footprint and the number of times it was covered, to sub-arcsecond precision.

\noindent $\bullet$ Then, in Section \ref{sec:full}, we explain how the combined information gathered in the previous section is condensed into `full' catalogs containing all unique entries per target and random that were reachable to a DESI fiber positioner, how veto masks can be applied to these full catalogs to restrict them to areas within which the selection function can be modelled, and the completeness statistics we determine for each target included in these catalogs. 

\noindent $\bullet$ Next, in Section \ref{sec:weights}, we describe how the selection function details such as completeness variations, imaging systematics in the target samples, and variation in the redshift success rate can be modelled, using the full catalogs. We then, in Section \ref{sec:null}, briefly discuss how the resulting catalogs allow for null tests of the selection function modelling and how catalog-level blinding can be applied. 

\noindent $\bullet$ The final pieces of the LSS pipeline are described in Section \ref{sec:clustering}, where we describe the production of the `clustering' catalogs that are meant to be input to DESI clustering measurement pipelines. This includes cutting the data to secure redshifts within a desired range, assigning redshifts to the random catalogs, and calculating weights meant to optimally account for changes in the observed density, e.g., as a function of redshift and the number of overlapping tiles.
 
\noindent We provide a summary and concluding remarks in Section \ref{sec:con}. 

The clustering catalog products are meant to enable the simple calculation of clustering statistics. The detailed information on the pipeline process presented in this work and its various assumptions allows biases in the resulting clustering statistics to be predicted and modeled. However, the details of such modeling are presented elsewhere. The specific choices regarding the pipeline applied to the DR1 data and the resulting 2-point clustering measurements\footnote{This includes various window functions required to translate between true and observed redshift-space clustering measurements.} are presented in \cite{DESI2024.II.KP3}. The ability to extract unbiased cosmological information from DESI LSS catalogs obtained from realistic simulations is the subject of many DESI DR1 studies, including \cite{KP4s4-Paillas,KP3s5-Pinon,KP5s6-Zhao}.  
 
 For convenience, a glossary defining relevant DESI terms, including those defined during the LSS pipeline process, is presented in an appendix.

\section{The Dark Energy Spectroscopic Instrument}
\label{sec:instrument}
The Dark Energy Spectroscopic Instrument (DESI \cite{DESIinstrument}) is installed on the Nicholas U. Mayall Telescope at Kitt Peak National Observatory in Arizona. It collects spectra using 5,020 robotically controlled fiber positioners \cite{DESIfocalplane} to move the same number of optical fibers onto pre-determined `target' objects on the sky \cite{DESItarget}. The light from 5,000 fibers\footnote{The light from a fixed set 20 of the fibers is sent to a camera that monitors the sky brightness \cite{DESIinstrument}} is divided into 10 equal-sized wedge-shaped regions known as `petals'. The light from the 500 fibers on each petal is carried to its own climate-controlled spectrograph (and thus there are 10 DESI spectrographs).
Within each spectrograph, the light is split into three wavelength bands, $B$ (3600-5800 \r{A}), $R$ (5760-7620 \r{A}), and $Z$ (7520-9824 \r{A}), each readout with one CCD detector.

\subsection{Focal Plane Geometry}
\begin{figure*}
    \centering 
    \includegraphics[width=0.5\columnwidth]{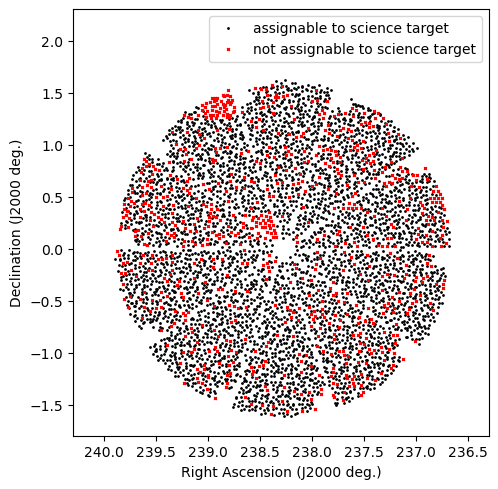}
    \includegraphics[width=0.47\columnwidth]{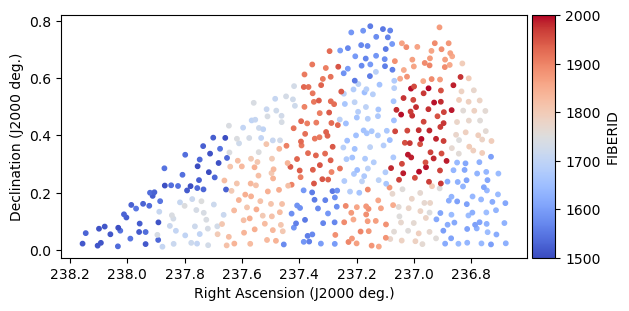}
   \caption{The left-hand plot displays the positions on the sky of each of the 5,000 fiber positioners that collect light to be sent to the spectrographs for DESI \texttt{TILEID} 1000. The 4,201 locations displayed in black were assigned to positioners that were enabled to be assigned science targets. The 799 in red were not and such areas for this tile are later masked from the LSS analysis. The right-hand panel displays the same positions, but for only one petal and colors show the \texttt{FIBERID}. The 500 \texttt{FIBERID} per petal have their light output sequentially across CCD detectors. }
    \label{fig:tilegeo}
\end{figure*}

The DESI focal plane subtends a radius of 1.6 degrees on the sky. However, within this area, there are many regions where no fibers can be placed. Fig. \ref{fig:tilegeo} shows the sky coordinates of assigned targets for one DESI `tile'. Each tile represents a specific pointing in celestial coordinates, with an associated set of targets assigned to fiber positioners. One can observe each of the ten petals, with a gap between each where no targets are assigned. There is a hole in the center of the focal plane where no positioners can reach. Finally, on each petal, there is a rectangular area missing at one edge of the outer radius. At this location on each, there is a `guide and focus array' camera, which observes stars to, as the name implies, keep the telescope on target and in focus. 

The pattern seen in the left-hand panel of Fig. \ref{fig:tilegeo} represents a fundamental building block of the geometry the DESI survey covers. However, the focal plane can be further divided into many sections based on the relevant hardware components. Individual components can fail but leave most of the instrument operational, with the effect simply being a small hole that needs to be masked in the LSS analysis. The greatest such region is a petal, while the smallest is a single fiber positioner. The operational status of each fiber positioner can change even between exposures taken on any given night. However, the positioner hardware database reflecting their status is updated at most daily. The assignments for a given night use this database to determine which positioners can potentially be assigned to science targets. The red points on the left-hand panel of Fig. \ref{fig:tilegeo} show fiber positioners that were flagged as not assignable to science targets on the night the tile was first observed, due to failures during calibration tests or other known hardware failures\footnote{When possible, these are used to obtain sky spectra.}.
 The contiguous blocks in the upper-left-hand petal were non-operational due to communication issues in the controller electronics, which were later fixed. Given the (at most) daily updates of the positioner hardware database\footnote{See section 5.13 of \cite{surveyops} for more details.}, positioners can only be lost during a night. However, the status of each positioner is determined and recorded in the metadata for every exposure, which allows any failures to be masked from the analysis in post-processing.
 
The light from the fibers is read out with a greater \texttt{FIBERID} number corresponding to a greater number in the CCD column \cite{DESIpipe}, thus, any issues with the CCDs are likely to remove a block of data that is contiguous in fiber number. Each \texttt{FIBERID} corresponds to a particular area on the focal plane, which is shown in the right-hand panel of Fig. \ref{fig:tilegeo}. One can observe a checker-board pattern, which serves to de-correlate the position on the CCD with the focal plane radius. A failure of only one of the $B, R, Z$ cameras will cause all of the data to be masked from LSS analysis; we require full wavelength coverage for `good hardware'. A common issue that DESI has faced is failures of particular amplifiers of the CCDs. Each amplifier\footnote{Readout of the DESI CCDs is possible in both two and four amplifier modes, but one faulty amplifier will always affect a full half of the spectra on the CCD.} is associated with half of the spectra on the CCD and thus when one fails, we mask half of the fibers on the petal (1 through 250 or 251 through 500). 

The combination of potential hardware effects described above means that the precise geometry of any two DESI tiles is unlikely to match. The angular patterns in any particular tile footprint should, however, look similar to those seen in Fig. \ref{fig:tilegeo}.

\subsection{Survey Strategy}

DESI survey operations are described in detail in \cite{surveyops}. Here, we detail how key aspects impact the creation of LSS catalogs. DESI dynamically divides its observing time into separate `bright' time and `dark' time programs, depending on observing conditions, with different target classes observed in each. DESI obtains targets for spectroscopic observations based on existing photometry \cite{DESItarget}. The different classes of targets are treated separately in dark and bright time. Thus any target can be observed in both dark and bright time\footnote{Though, this would mean it satisfied the target selection cuts of both a bright time and a dark time sample.}; one such observation will not influence the other. For extra-Galactic targets, observations of the Bright Galaxy Sample \cite{BGStarget} happen in bright time. The Luminous Red Galaxies (LRGs \cite{LRGtarget}), quasars (QSO \cite{QSOtarget}), and Emission Line Galaxies (ELGs \cite{ELGtarget}) are observed in dark time. 

Observed data from every night are processed by the DESI spectroscopic pipeline \cite{DESIpipe} each morning following observations for immediate quality checks. The results are referred to as the `daily' products. Tiles that have been observed with a cumulative effective observing time that is at least 85\% of the target threshold are considered for approval, via a quality assurance (QA) program that inspects, e.g., the redshift success rates, ${\rm d}N/{\rm d}z$, and the residuals of sky fibers following sky subtraction\footnote{Section 5.10 of \cite{surveyops} has further details.}. 

The information on the targets observed on the accepted tiles is fed back into the `merged target ledger' (MTL), which contains all of the details on DESI targets necessary to pass them to fiber assignment. The MTL contains a ledger that details the observing history of all targets, including the initial unobserved state for all. Separate ledgers exist for dark and bright times. The targets that were observed with properly working hardware have their state updated. For galaxies and quasars with redshifts $<2.1$, this means their priorities are reduced such that they will only be observed again if no other science targets are available to a given fiber positioner. Quasars with $z>2.1$ instead get up to a total of four observations, until their priorities are similarly reduced. This is to increase the signal-to-noise of the Lyman-$\alpha$ forest absorption features in the spectra. Targets that were attempted to be observed with poorly performing (`bad') hardware are instead masked from updates, i.e., they are treated as if the observation did not happen. The flags used to mask bad hardware are detailed in section 6.3.1 of \cite{surveyops} and are the same as we describe in Section \ref{sec:specinfo}, which ultimately allows us to apply the same mask to data and randoms, at a resolution of individual fiber positioners.

To observe all DESI target classes to a high completeness and without significant holes, each area on the sky is covered by multiple tiles. The expected distribution for the number of overlapping tiles for completed dark and bright time surveys is shown in figure 2 of \cite{surveyops}, and on average it is greater than 5 for dark time and 3 for bright time. This assumes 7 passes of non-overlapping tiles in dark time and 4 in bright time. Additional passes can be added, e.g., if observations are more efficient than originally predicted and extra time is available. DESI prioritizes the completion of observations at a given area on the sky, as opposed to covering the full footprint. However, based on the MTL process described above, any observed tile cannot have overlapping observations until it has passed QA. The QA process was still being fully defined during the first two months of DESI main survey operations. Thus, the data from this period, used for the first large-scale clustering studies presented in \cite{Moon23}, consists primarily of non-overlapping tiles. The resulting DR1 coverage is thus a mix of area covered with a large and small number of overlapping tiles, see \cite{DESI2024.II.KP3}.

\section{Gathering and Processing Inputs}
\label{sec:inputs}
DESI targets are selected based on photometry from Data Release 9 of Legacy Survey \cite{LS.Overview.Dey.2019,LS.dr9.Schegel.2024} imaging \cite{DESItarget}. DESI LSS catalogs are divided into separate tracer catalogs based on their designation at the time of targeting, as defined by \textsc{desitarget} \citep{DESItarget}. This means, for example, that an LRG target determined to be a star-forming galaxy based on its spectrum will always remain an LRG in the LSS catalogs. Observations that do not yield secure redshifts or are outside of the desired redshift range are eventually discarded, but not until the production of the `clustering' catalogs described in Section \ref{sec:clustering}. After defining samples based on their targeting, sub-samples can be defined based on any combination of photometric and spectroscopic properties. This will be discussed in Section \ref{sec:subsamples}. Throughout, we will refer to a `target' as an object that has been selected for spectroscopic follow-up but has not necessarily been observed. A qualifier will always specify its state, where relevant (e.g., `observed', `reachable').
 
 For the data, the process described in the following subsections is repeated for each of the tracer types. We first select all of the DESI targets for a given type and include only a sub-selection of the columns relevant to LSS analysis. We save the files to disk with a root naming scheme that is based on the \texttt{targetmask} function in \textsc{desitarget} (see \cite{DESItarget} for more details) and will be shared by the rest of the catalog products. For the DR1 analysis, five samples are produced: \texttt{LRG}, \texttt{ELG\_LOPnotqso}, \texttt{QSO}, \texttt{BGS\_ANY}, and \texttt{BGS\_BRIGHT}. The LRG and QSO samples are selected purely based on their \texttt{DESI\_TARGET} bit value and passing their name to the \texttt{targetmask} function. For the \texttt{ELG\_LOPnotqso} sample, targets satisfying the \texttt{ELG\_LOP} \texttt{DESI\_TARGET} bit are kept and those also satisfying the QSO \texttt{DESI\_TARGET} bit are rejected (these are considered only in the QSO catalogs). The \texttt{BGS\_BRIGHT} sample is produced by selecting based on the \texttt{BGS\_TARGET} value (instead of \texttt{DESI\_TARGET}) and it is a subset of the \texttt{BGS\_ANY} sample. Dark time targets (LRG, QSO, ELG) are compiled only for dark time tiles, and bright time targets (BGS) for bright time tiles.

The ELG naming convention can be explained as follows (see \cite{ELGtarget,DESItarget} for the full details): The ELG sample overall is split into a higher priority (designated \texttt{ELG\_LOP}) and lower priority sample (designated \texttt{ELG\_VLO}), with the higher priority sample comprising ~75\% of the targets and containing most that have $z>1.1$. The designation \texttt{ELG\_HIP} is reserved for the 10\% of ELG targets (with no distinction between \texttt{ELG\_LOP/VLO}) that are randomly selected to have the same assignment priority as LRG targets. The ELG targets are given this targeting bit in addition to their original bit (i.e., selecting \texttt{ELG\_LOP} will always select the full original \texttt{ELG\_LOP} sample defined via photometric cuts).

Catalogs can be constructed for the full \texttt{ELG(notqso)}. We have not done so for DR1 as we have instead focused on the higher assignment priority (and thus more complete) samples. For simplicity, the \texttt{ELG\_LOPnotqso} sample is referred to simply as `ELG' in all DR1 analyses. Future analyses are likely to include the lower-priority ELG sample in cosmological analyses.

We also use the randoms produced by \textsc{desitarget} \citep{DESItarget}. These are random points in celestial coordinates isotropically occupying the area covered by Legacy Survey \cite{LS.Overview.Dey.2019,LS.dr9.Schegel.2024} imaging. They are divided into files that each have a density of 2500/deg$^2$ and a total of 200 are available. As of DR1, the DESI LSS pipeline uses 18 of them.
 The 2500/deg$^2$ is convenient, as it allows for quick determination of footprint area after various cuts. In the following subsections, the randoms are processed separately only for dark and bright time; separate processing for randoms based on the particular tracer type does not happen until later steps.

The details of the above discussion are independent of actual DESI observations. To produce LSS catalogs, we require two fundamental pieces of information based on the DESI observations:\begin{itemize}
    \item 
 1) The area on the sky where we could have observed (i.e., the `footprint'), based on where the telescope was pointed and hardware specifications. 
 \item 
 2) Redshift determinations and associated metadata extracted from the observed spectra
\end{itemize}
The process of obtaining this information is described in the following subsections.

\subsection{Collecting Reachable Targets}
\label{sec:combinedtargets}
We use the DESI \textsc{fiberassign} \cite{FBA.Raichoor.2024} software\footnote{\url{https://github.com/desihub/fiberassign} } to define the DESI footprint and collect the information on all targets within it. The properties of all cases in which a DESI target is reachable to a given fiber positioner are determined for every tile; each case represents a `potential assignment'. Collecting this information is required to determine completeness statistics and correct for their variations properly. Each fiber positioner is expected to be able to reach many DESI targets. All of these will be potential assignments, but only one will receive the actual assignment, which is given to the target that has the highest \texttt{PRIORITY} (and then \texttt{SUBPRIORITY} if multiple have the same \texttt{PRIORITY}). The potential assignments information is recorded within the same file that designates the final (actual) fiber positioner assignments\footnote{within the \texttt{POTENTIAL_ASSIGNMENTS} HDU, for the files described here: \url{https://desidatamodel.readthedocs.io/en/latest/DESI_TARGET/fiberassign/tiles/TILES_VERSION/TILEXX/uncompressed-fiberassign-TILEID.html}}. 

For a given set of tiles (e.g., those included in DR1), we concatenate the potential assignments information for each of the target samples defined above. When doing so, we match (via \texttt{TARGETID}) to the target information at the time of assignment (recorded in the file header of the \textsc{fiberassign} file for each tile), as recorded in the MTL, thereby ensuring we keep all relevant metadata. The information on all potential assignments on all tiles and fiber positioners is kept, without any consideration of whether or when an actual assignment occurred. Thus, each \texttt{TARGETID} is likely to appear multiple times; at least once for every tile containing a fiber positioner that could reach it and potentially multiple times for each such tile (given the patrol radii for fiber positioners overlap). The final result is the information, for a given set of DESI tiles, on all of the cases for which targets of the given target sample were reachable by a DESI fiber positioner; this will vastly outnumber the observed cases. This thus forms a super-set of the targets within the observed DESI footprint, which we write to disk and then use in subsequent steps. We denote these catalogs as the `combined target information catalogs'\footnote{Their location in the data model is here: \url{https://desidatamodel.readthedocs.io/en/latest/DESI_ROOT/vac/RELEASE/lss/VERSION/potential_assignments/data/index.html}. The main update for DR1 is that the data catalogs are split by tracer type instead of program.}.

The \textsc{fiberassign} software can be run efficiently on random samples to obtain only their potential assignments, using the specific hardware state and observing information as for the data. The information required to reproduce the data assignment is stored for every tile and verified based on automated software to exactly reproduce the original assignments\footnote {This is non-trivial because different computing systems produce each version of the assignments and passing requires that the hardware database is properly synced between the software environments.} before DESI survey operations accepts the information observed on the tile. Running the potential assignments software using the same settings but applying it to the random samples thus finds all of the reachable randoms for the same set of tiles. At this stage, the randoms do not need to be separated by tracer sample; the information will be the same for the same set of tiles. Thus, the combined target information catalogs for each of the (18 for DR1) randoms we process are saved in separate dark and bright time files (since dark and bright time use different sets of tiles)\footnote{Their location in the data model is here: \url{https://desidatamodel.readthedocs.io/en/latest/DESI_ROOT/vac/RELEASE/lss/VERSION/potential_assignments/random/index.html}.}.

The potential assignment information for data and randoms can be gathered immediately following any DESI observations. We therefore do so on a roughly bi-weekly basis and thus keep up to date with the survey progress. This allows any particular issues to be identified and solved before the data release spectroscopic processing and then allows a fast turn-around on the production and validation of LSS catalog products after the data release processing.

A caveat to the above discussion is that the compiled information is a super-set of the targets that were actually reachable. It does not incorporate regions that were excluded, e.g., for being too close to the edge of a petal. These are referred to in the \textsc{fiberassign} software as `collisions'. When obtaining the assignments for a tile, the \textsc{fiberassign} software first determines all potential assignments, ignoring any collisions. The collisions are only rejected when determining the actual assignments and the potential assignments recorded in the \textsc{fiberassign} files thus include the collisions. To properly define the footprint, we thus run a function from the \textsc{fiberassign} code, developed to address this issue, a posteriori to reject the approximately 5\% of targets and randoms that were not truly reachable. For the data, these collisions are removed at the beginning of the steps to create the `full' catalogs described in Section \ref{sec:full}. For the randoms, we record the collision information at the same time as running the software that determines the potential assignments, as running these together is more efficient than running them in separate steps.

\subsection{Collecting Spectroscopic Information}
\label{sec:specinfo}
We collect information from the spectroscopic reduction and redshift fitting pipeline both from the `daily' version and for the version associated with the data release. For DR1, homogeneous processing applied to the full dataset is referred to as `Iron'. This includes data taken from May 14, 2021 through to June 14, 2022. In the Iron release, the majority of the spectroscopic information we use has already been concatenated across the observed tiles by the DESI spectroscopic pipeline team and released as the per-tile cumulative redshift catalogs\footnote{\url{https://desidatamodel.readthedocs.io/en/latest/DESI_SPECTRO_REDUX/SPECPROD/zcatalog/zall-tilecumulative-SPECPROD.html}}. 

The daily version of the spectroscopic pipeline is run as soon after data collection as possible, typically the following morning. No official redshift catalog is produced by the spectroscopic pipeline based on the daily data. Thus, for processing the daily spectroscopic reductions through the LSS pipeline, we instead concatenate the information produced for each observed tile, on the same roughly bi-weekly cadence as we obtain the combined target information catalogs. This ultimately allows LSS catalogs to be produced that are up to date with the survey progress. 

For both the daily and data release spectroscopic products, as part of the LSS catalog process, we concatenate three additional classes of information across the given set of tiles:\begin{itemize}
    \item The `zmtl' files, from which the key piece of information for our purposes is whether the observation would have caused an update to the MTL ledger, based on the available information on the hardware performance. This information is later used to determine the data and randoms associated with `good' hardware.
    \item The files containing information on emission line flux. This information is later used to determine whether an ELG observation yields a good redshift.
    \item QSO catalogs, per tile, the data that satisfies the criteria determining whether an observation results in a secure QSO is kept. This information is later used to determine whether a QSO observation yields a good redshift.
\end{itemize}
The concatenated `zmtl' files are joined with the redshift catalog, matching based on \texttt{TARGETID} and \texttt{TILEID}. Both the redshift and zmtl files have a column named \texttt{ZWARN}, but the zmtl files contain extra bits that encode good hardware information. Thus, we convert the zmtl column name to \texttt{ZWARN\_MTL}. This produces the `combined spectroscopic information catalog'.

Information from the combined spectroscopic information catalog is joined to both the data and random combined target information catalogs. We first cut the information to the list of \texttt{TILEID} to be included for any data release. For the data, we then perform a left-outer join\footnote{Such a join will keep all of the rows of the table that is being matched to.} so that all targets are kept, regardless of whether they were observed. The join is performed by matching on \texttt{TARGETID},\texttt{FIBER},\texttt{TILEID} and thus properly matches observed targets to their tile and fiber. For the randoms, we match only to the \texttt{FIBER} and \texttt{TILEID}; thus a match should exist for all. For convenience, we combine the \texttt{TILEID} and \texttt{LOCATION} columns to create a unique identifier \texttt{TILELOCID} for the observation (\texttt{LOCATION} is the positioner ID and thus there is a one-to-one mapping between values of \texttt{FIBER} and \texttt{LOCATION}) via
\begin{equation}
     \texttt{TILELOCID} = 10000\texttt{TILEID} +\texttt{LOCATION}.
\end{equation}
All data (including randoms) are given a \texttt{TILELOCID}, as we are only considering data that was identified as a potential assignment and can thus be associated with a given DESI tile and fiber. The \texttt{TILELOCID} allows all of the data (again, including randoms) to be linked to the observational metadata, including the targets that were not observed. 

The resulting data and random catalogs are saved to disk\footnote{In the data model here \url{https://desidatamodel.readthedocs.io/en/latest/DESI_ROOT/vac/RELEASE/lss/VERSION/inputs_wspec/index.html}. The main update for DR1 is that the data catalogs are divided by tracer type instead of program. } and used as inputs for the next step the LSS catalog pipeline, described in Section \ref{sec:full}. We will refer to them as the `combined target and spectroscopic information catalogs'.

\subsection{Counting Target Appearances}
\label{sec:count}
The catalogs described above are used to determine information on all of the times a target appeared on a fiber that was observed with ``good hardware". For the good hardware definition, we use the same $H_{\rm gd}$ quantity defined in Section \ref{sec:vetos}. We obtain the information on one entry per \texttt{TARGETID} and determine three new quantities, which are recorded in the following columns in the `full' catalogs described in Section \ref{sec:full}:

\textbullet~ \texttt{NTILE} An integer denoting the number of tiles the target appeared on; we will denote this $N_{\rm tile}$ in equations.

\textbullet~ \texttt{TILES} A string listing the tiles that the target appeared on, using the \texttt{TILEID}s sorted in ascending order and separated by `-'. Each unique \texttt{TILES} represents a unique group of overlapping tiles that we will use later to group the data and refer to as $t_{\rm group}$.

\textbullet~ \texttt{TILELOCIDS} A string listing the \texttt{TILELOCID}s separated by '-'.

\begin{figure*}
    \centering 
    \includegraphics[width=0.47\columnwidth]{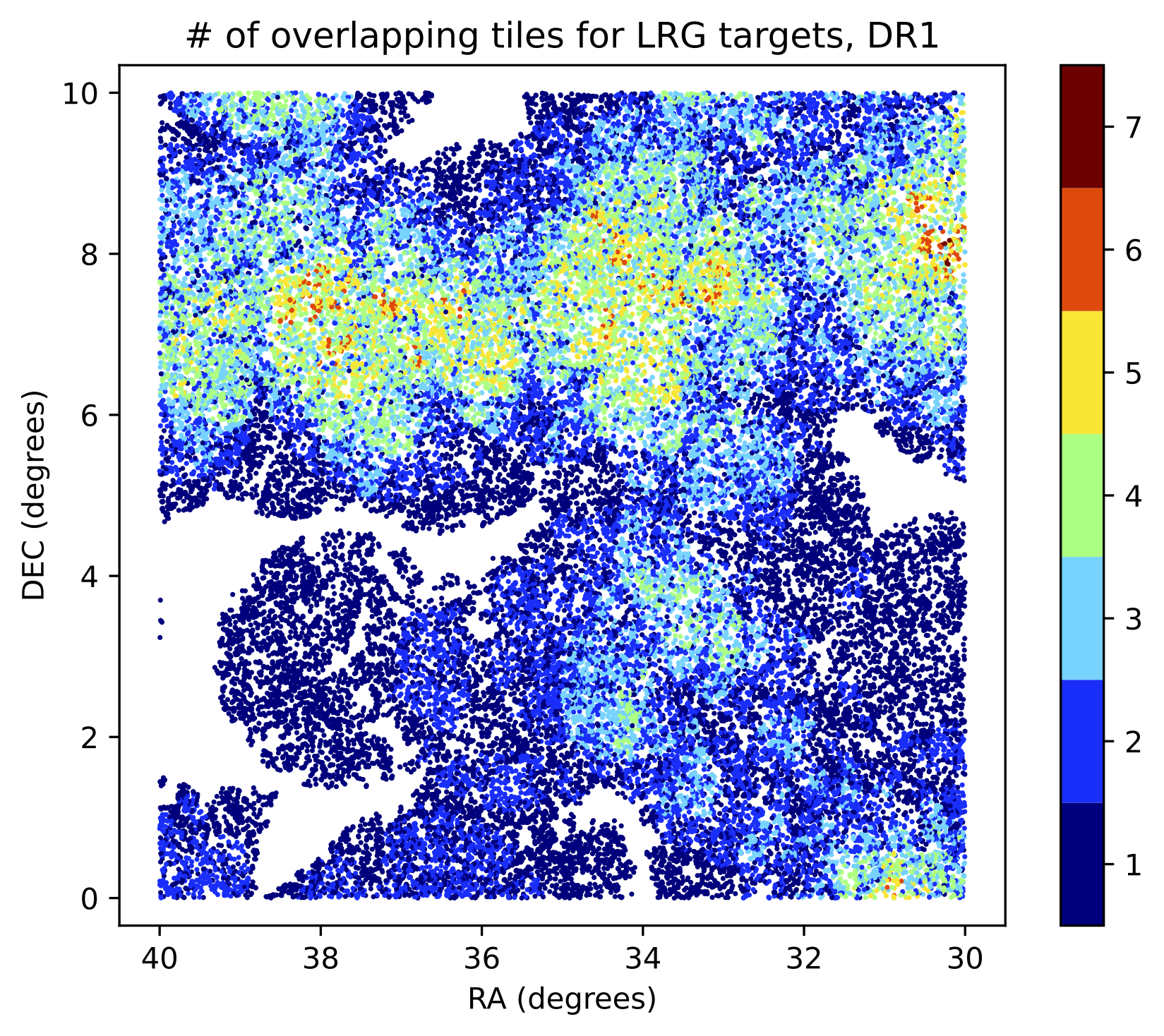}
    \includegraphics[width=0.47\columnwidth]{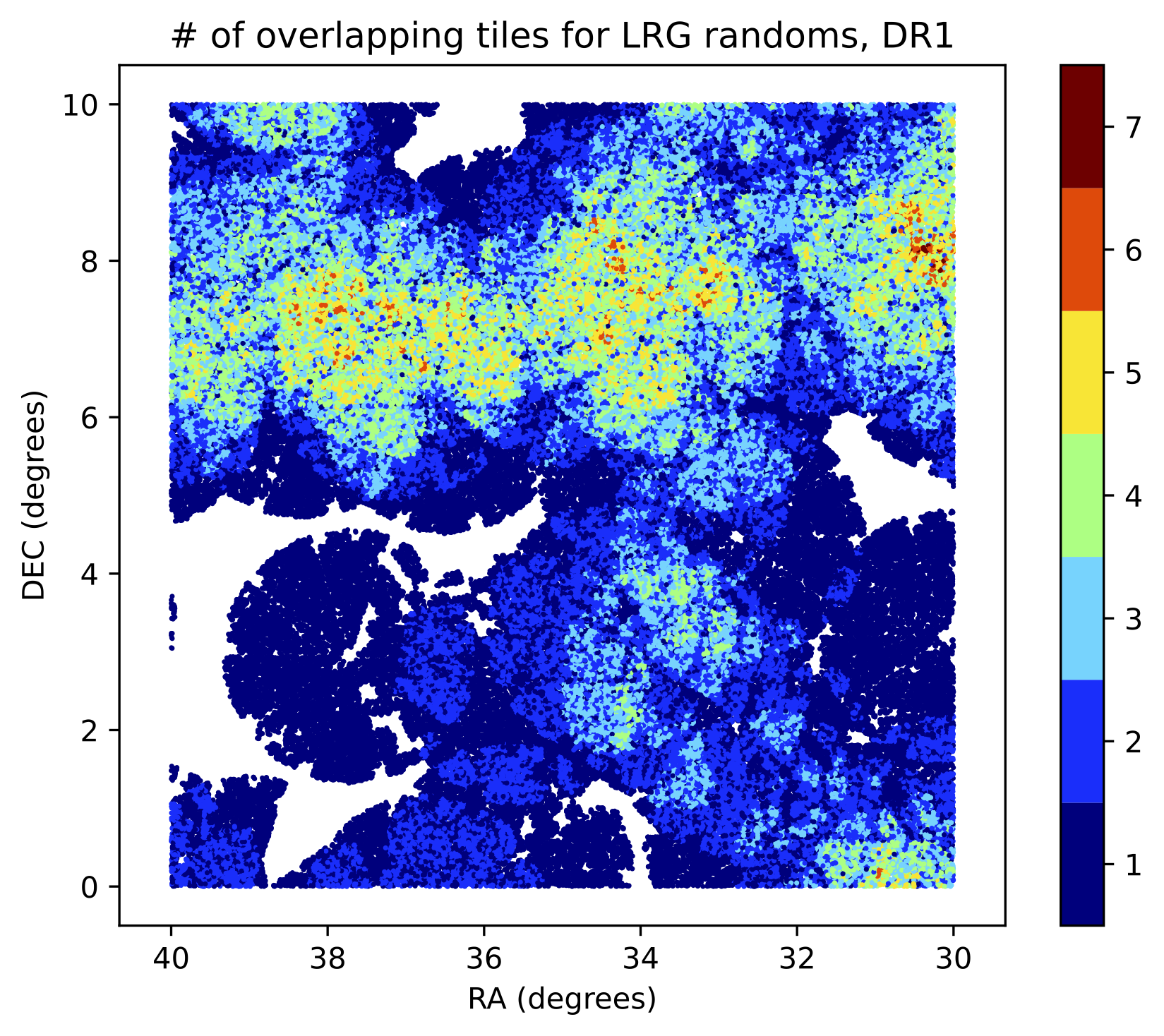}
    \includegraphics[width=0.47\columnwidth]{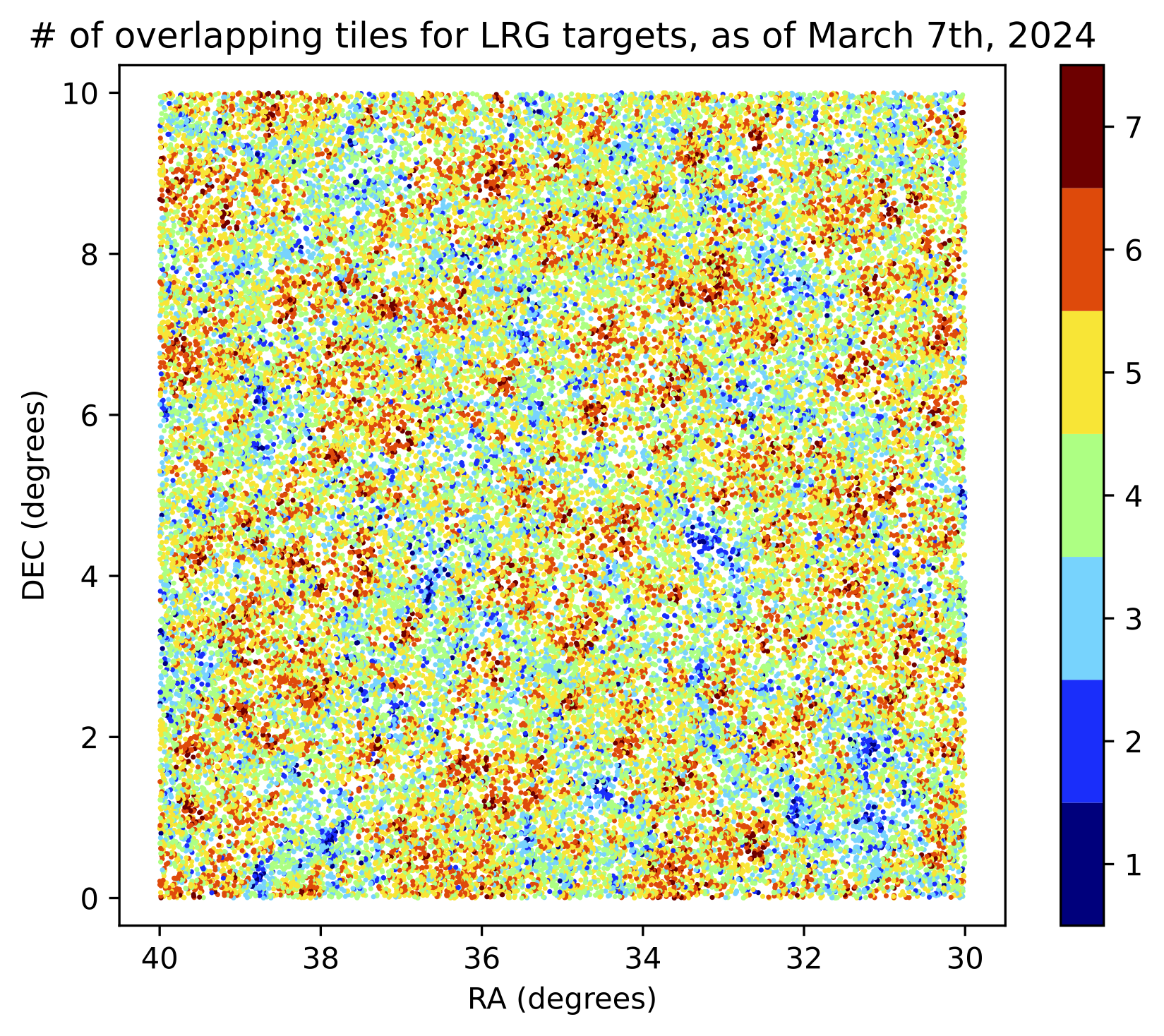}
    \includegraphics[width=0.47\columnwidth]{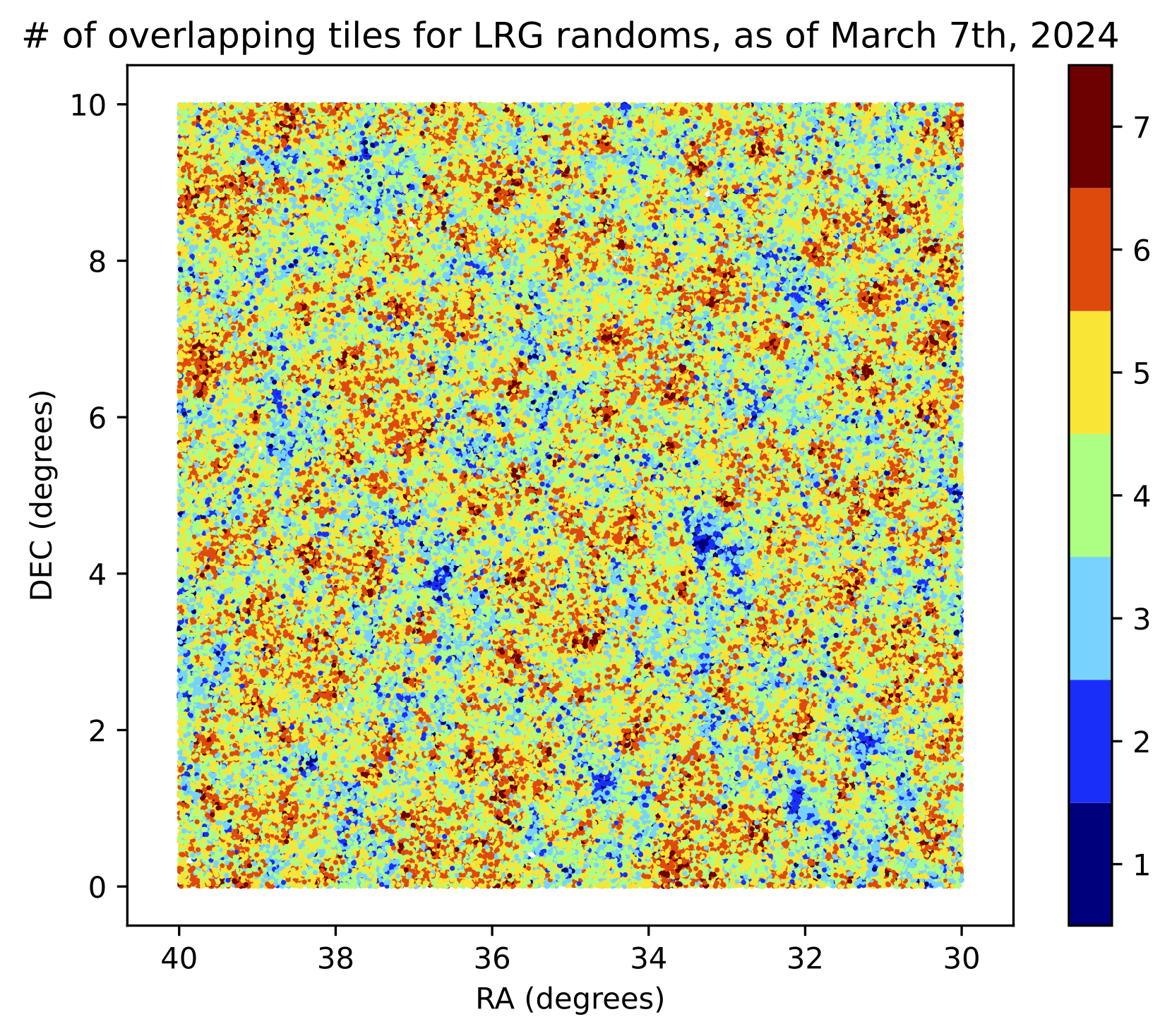}
   \caption{Maps of the number of overlapping tiles, $n_{\rm TILE}$ in a small selection of the DESI area: The top panels show results within the DR1 (May 14, 2021 through to June 14, 2022) footprint and the bottom panels show the results as of March 7th, 2024. The left-hand panels show $n_{\rm TILE}$ at the location of potential LRG assignments. The right-hand panel shows the same information assigned to the random sample. One can observe they trace the same small-scale structure. }
    \label{fig:NTILE}
\end{figure*} 

The process described above provides the basic geometrical information on the area of the sky that DESI has observed, for both data and randoms. The resolution is limited only by the accuracy at which the DESI \textsc{fiberassign} software can properly predict a target will be reachable by the DESI focal plane system. The mean difference between the predicted and final focal position for successful observations in DR1 dark time is 0.04mm, which corresponds to less than 1 arcsecond on the sky. The small-scale geometry of the DESI footprint is illustrated in Fig. \ref{fig:NTILE}, where a small selection of the DR1 data is plotted for LRG targets and randoms in the top panels. All targets that were reachable by good hardware are shown. Within this region, the number of overlapping tiles varies between 1 and 7. In areas with $n_{\rm TILE}>3$, one can see that exact numbers vary on small scales but are the same for both targets and randoms.

In the bottom panels of Fig. \ref{fig:NTILE}, the coverage as of March 7th, 2024 is shown. These results are obtained from the daily version of the LSS catalog pipeline. One can observe that the coverage has improved substantially, with the entire area covered at least once and most of it more than four times. Small pockets exist with only one tile coverage, and the completeness will be lower in those areas. The displayed mapping will be used for the completeness corrections.
 For LRG targets, one observes some areas that are still white, which are simply areas with no LRG targets (at the resolution of the figure), given we are plotting all targets that were reachable by good hardware.

\section{Target Catalogs in DESI Footprint}
\label{sec:full}
In the previous section, we detailed how the information associated with each case in which a target was reachable is gathered and stored in combined target and spectroscopic information catalogs, typically resulting in many entries per \texttt{TARGETID}. The next step in the DESI LSS pipeline is to produce the `full' catalogs that contain entries for all reachable targets, whether or not they were observed, but are cut to unique \texttt{TARGETID}. These catalogs are produced per supported tracer type (described in \ref{sec:inputs}). We describe the process of creating these catalogs and the stages they go through in this section. Weights to account for fluctuations in the selection function are eventually added to these catalogs but will described in Section \ref{sec:weights}.

\subsection{Full Catalogs, Without Vetos}
\label{sec:full_noveto}
The `full noveto' catalogs are constructed starting from the joined combined target and spectroscopic information catalogs (which have already been divided by target type). We wish to cut to unique \texttt{TARGETID}. We cannot do so randomly, as, e.g., we must keep the cases that have been observed. Similar to SV (\cite{EDR}; but with some important details different), we define three Boolean quantities that each depend on the particular tile and fiber. Many targets will be potential assignments for the same tile and fiber (with only one possibly observed). Thus, the designated quantities are likely to be shared by many targets. They are:

\textbullet~ $H_{\rm gd}$: The particular fiber on the given tile was observed with good hardware. Values with \texttt{false} will later be used as a veto. We provide further details on how each $H_{\rm gd}$ determination is made in Section \ref{sec:vetos} (the subscript $_{\rm gd}$ is short for `good').

\textbullet~ $T_{\rm fa}$: The particular combination of fiber and tile (but not necessarily target) was assigned and observed; i.e., for the LRG sample, this quantity will be \texttt{True} if any LRG was assigned and observed on the given fiber and tile, but, given that we have already cut to LRG targets at this point, it will be \texttt{False} if any other type of target was assigned to the particular combination of fiber and tile (the subscript $_{\rm fa}$ stands for `fiber assigned').

\textbullet~ $P_{\rm gd}$: The \texttt{PRIORITY} of the target that was assigned at the given tile and fiber was less than or equal to the maximum \texttt{PRIORITY} allowed for the given target class. We describe how these values are set for DR1 in the text that follows. This will later be used as a veto, and further details of its definition are given in Section \ref{sec:vetos}.

\vspace{4pt}
\noindent A fourth quantity is defined that depends on the particular target (in addition to the fiber and tile):

\textbullet~ $L_{\rm fa}$: The particular target, fiber, and tile was assigned and observed.

These quantities are combined to create a value to sort by
\begin{equation}
    v_{\rm sort} =  L_{\rm fa}  H_{\rm gd} P_{\rm gd} Q_{\rm pri}+T_{\rm fa} H_{\rm gd} P_{\rm gd}+H_{\rm gd} + P_{\rm gd},
\end{equation}
where $Q_{\rm pri}$ is 1 for all tracers except for QSO and for QSO it is the \textsc{priority} value; this prioritizes the first observations of QSO.\footnote{The first QSO observation is used to determine whether it will get three more observations at high \texttt{PRIORITY} and thus only the first observation provides a fair test for determining redshift success rates (as the likelihood of success on a subsequent observation is conditioned on the success of the first observation).} We sort by this $v_{\rm sort}$ in ascending order and then cut to unique targets by selecting the last entry for each unique \texttt{TARGETID}. 

The purpose of the sort is to 
\begin{itemize}
    \item ensure that targets observed with good hardware are kept,
    \item ensure that targets with \texttt{TILELOCID} shared by an observed target of the same type are kept ahead of those that are not,
    \item and ensure that any entries not in a veto mask are prioritized relative to those that are.
\end{itemize}
This maximizes both the number of observed targets and the number of targets associated with an observation of the given type; this ultimately maximizes the $f_{\rm tile}$ completeness defined in Section \ref{sec:comp} and Eq. \ref{eq:ftile}. Table \ref{tab:sortv} shows values of the above boolean columns for a particular LRG target that was reachable on multiple tiles. The $v_{\rm sort}$ is greatest for the case in which the target was actually observed, ensuring that is the entry kept. The next greatest $v_{\rm sort}$ value (3) is for the case where the \texttt{TILELOCID} is the same as for a separate LRG target that was observed. We note that, as of DR1, it is possible for, e.g., the maximum $v_{\rm sort}$ value of a target to be 3 and for that value to occur at multiple \texttt{TILELOCID}. In such a situation, the \texttt{TILELOCID} kept will depend on the ordering before the sort. 

\begin{table*}
\centering
\caption{\label{tab:sortv} Example values that contribute to how the data on reachable targets is sorted before cutting to unique \texttt{TARGETID}, for a particular LRG target. The \texttt{LOCATION} value corresponds to a unique fiber positioner.}
\begin{tabular}{cccccccc}
\hline\hline
\texttt{TARGETID} & \texttt{TILEID} & \texttt{LOCATION} & $L_{\rm fa}$ & $T_{\rm fa}$ & $H_{\rm gd}$ & $P_{\rm gd}$ & $v_{\rm sort}$\\
39627569254897060 & 2480 & 8337 & 0 & 0 & 1 & 1 &2\\
39627569254897060 & 5609 & 4091 & 0 & 1 & 1 & 1 & 3\\
39627569254897060 & 11875 & 418 & 0 & 0 & 0 & 1 & 1\\
39627569254897060 & 11875 & 419 & 1 & 1 & 1 & 1 & 4\\
\end{tabular}
\end{table*}

Before the sort, we obtain the footprint coverage information described in Section \ref{sec:count}. For DR1, we do this simultaneously, to ensure full consistency in the $H_{\rm gd}$ definition used. The coverage information is determined per target. Thus, after the data described above has been cut to unique targets, it is matched via \texttt{TARGETID} to the coverage information. In this way, in addition to providing the information on the number of overlapping tiles for each target, we retain the information on all tiles and \texttt{TILELOCID} that the given target was reachable on (where $H_{\rm gd}$ was \texttt{True}), encoded in the \texttt{TILES} and \texttt{TILELOCIDS} columns.

At this stage, we join to any extra information on the spectroscopic observations that was not included in the redshift catalog, e.g., the QSO catalog that was compiled separately. A left-outer join of the spectroscopic columns to the table containing all of the target information is performed matching on the \texttt{TARGETID} and \texttt{TILELOCID} columns. For ELGs, this includes information on the OII emission line fits, produced with the spectroscopic release (as of DR1, these are in separate catalogs from the redshift catalog already joined to, but this is likely to change for future releases). For QSO, this is the information from the separate QSO catalog. The result is the `full noveto' data LSS catalog\footnote{\url{https://desidatamodel.readthedocs.io/en/latest/DESI_ROOT/vac/RELEASE/lss/VERSION/LSScats/full/index.html}}. Notably, even after doing the sort and cutting to unique targets, some of the \texttt{TILELOCID} will not be associated with targets of the given type, i.e., the given \texttt{TILELOCID} could have been assigned a different type of tracer, sky, or a standard. The sort minimizes these cases, but this remains an important aspect of the completeness, discussed in Section \ref{sec:comp}.

For randoms, we must also cut to unique \texttt{TARGETID} and we do so separately for each tracer type, as the tracer information is included in the sort. The input for each is the dark or bright time random combined target and spectroscopic information catalog. We also apply the cuts to areas with at least one observation in the $g,r,z$ photometric bands, and the LegacySurvey DR9 imaging maskbits\footnote{See \url{https://www.legacysurvey.org/dr9/bitmasks/}; the data used for the number of exposures and maskbits are the {\tt nexp} and {\tt maskbits} files listed here: \url{https://www.legacysurvey.org/dr9/files/}} that were already applied to the target samples. Bits 1 and 13 were applied for bright time targets and additionally bit 12 for dark time targets. In addition to the boolean quantities used to sort the data, we define one additional quantity:

\textbullet~ $Z_{\rm poss}$: The combination of tile and fiber was either assigned to a target of the given target class or no unassigned targets of the given target class were reachable by the fiber on this tile. Its inclusion is a legacy of previous versions of the pipeline and does not currently serve any particular purpose (but also does no harm).

The randoms are then sorted by
\begin{equation}
    v_{\rm sort} = H_{\rm gd} P_{\rm gd} Z_{\rm poss} + H_{\rm gd} P_{\rm gd}.
\end{equation}
As for the data, the primary purpose of the sort is to ensure that any entries not in a veto mask are prioritized relative to those that are and we cut to unique random points by selecting the highest $v_{\rm sort}$ value for each. The result is the `full noveto' random LSS catalog.

\subsection{Applying Vetos}
\label{sec:vetos}
The `full noveto' catalogs have kept all of the data and random targets that were deemed reachable when \textsc{fiberassign} was run. The data and randoms at this point will have matching sky footprints and any angular clustering measurements should match that of the full target samples (taking the reduced footprint size and systematic targeting variations within the full footprint into account.). However, to cut the sample to the sky area where redshift could have been measured successfully and the target sample's selection function modelled, we apply a series of veto cuts.

We cut to \texttt{True} values of the two Boolean quantities, $H_{\rm gd}$ and $P_{\rm gd}$. $H_{\rm gd}$ being \texttt{True} denotes that observation corresponding to the particular tile and fiber used `good' hardware. This is primarily determined from the \texttt{ZWARN\_MTL} quantity defined in Section \ref{sec:specinfo}. The cuts applied based on the \texttt{ZWARN\_MTL} flag are described in section 6.3.1 of \cite{surveyops}. Notably, we use the version of the \texttt{ZWARN\_MTL} information obtained with the same spectroscopic reduction as the redshift information we use (e.g., iron for DR1). The \texttt{ZWARN\_MTL} can change with spectroscopic reduction and the information we use can thus be different from what was used for the MTL decisions. This is no problem for defining the footprint but does present a small complication for the `alternative-MTL' process \cite{KP3s7-Lasker} used to obtain joint probabilities of assignments, described in Section \ref{sec:comp}. 

We apply two additional criteria to define $H_{\rm gd}$. The first is that we flag a set of fibers as `bad', which we re-evaluate with each data release\footnote{In the future, we will likely consider different date ranges for different blocks of fibers, corresponding to distinct changes in the hardware, such as changing a CCD detector. }. The details of determining the bad fiber list for DR1 analysis are given in \cite{KP3s3-Krolewski}, where 60 are flagged and removed from the LSS catalogs. All \texttt{TILELOCID} (in data and randoms) associated with such fibers have $H_{\rm gd}$ set to \texttt{False}. The second is that we apply a threshold on the spectroscopic depth, quantified by the ‘template signal to noise ratio squared’ value, (\texttt{TSNR2\_(tracer)} in the catalogs), defined in \cite{DESIpipe}. This is determined per observation, using a constant template spectrum, and thus depends only on the effective observing time and the detector noise associated with the particular fiber (and it explicitly does not depend on the source brightness). For DR1, we apply a threshold of  \texttt{TSNR2\_ELG}$>80$ in dark time and \texttt{TSNR2\_BGS}$>1000$ in bright time. All \texttt{TILELOCID} (in data and randoms) below these thresholds have $H_{\rm gd}$ set to \texttt{False}, which are 1\% of the DR1 data for both dark and bright time. The particular choice of values will be re-evaluated with each data release, but was motivated by the fact that redshift success rates fall dramatically for data below these thresholds. The effect of the good hardware veto on the footprint will be the same for all tracer types observed in the given program (dark or bright).

We use the $P_{\rm gd}$ definition to remove data and randoms that could not have been assigned to particular tile and fiber combinations due to the priority of the target that was assigned. The choice of exactly what to apply here is not completely obvious and may change with each data release. Different tracer types overlap in redshift and 
are thus correlated. Thus, the structure of any mask determined based on the priority values of one LSS sample is likely to have at least a small level of correlation with the true structure of other LSS samples.\footnote{Further discussion and analysis of the issue can be found in \cite{DESI2024.II.KP3} and references therein.} For DR1 analyses, we remove any QSO targets from the ELG sample and use the same priority value threshold of 3200 for the ELG and LRG samples (recall that \texttt{ELG\_HIP} have their priority boosted to 3200, which is the same as LRG). In most cases, LRG targets have higher priority than ELG targets. However, given that LRGs have a strong angular clustering signal, the LRG and ELG overlap highly in redshift, and 10\% of ELG have the same priority as LRG, it is undesirable to define any veto mask for ELG based on LRG. For QSO, anything assigned at a priority greater than 3400 is vetoed. For DR1, this only includes a small number of strong lens candidates that had priority 4000. For BGS, anything with a priority greater than 2100 is vetoed; only White Dwarf candidates (priority 2998) and strong lens candidates have such high priorities for bright time observations. The effect of the priority mask on the footprint will thus generally depend on the tracer, though for DR1 analysis, it is the same for ELG and LRG.

In general, we will apply additional vetos based on the imaging data. These may come from the LegacySurvey DR9 imaging maskbits\footnote{https://www.legacysurvey.org/dr9/bitmasks/}, or from more carefully considering the details of a particular tracer, as done for LRGs in \cite{LRGtarget}. When using such a mask for a particular tracer, the mask values are added to the data and random catalogs, named like, e.g., \texttt{LRG_MASK}. The mask will in general be different for different tracer types and is likely to be refined with each data release. For DR1, the choices and their impact are described in \cite{DESI2024.II.KP3}. After applying all three of these vetos, the catalogs are saved without `noveto' in the name. 

\begin{figure*}
    \centering 
    \includegraphics[width=0.47\columnwidth]{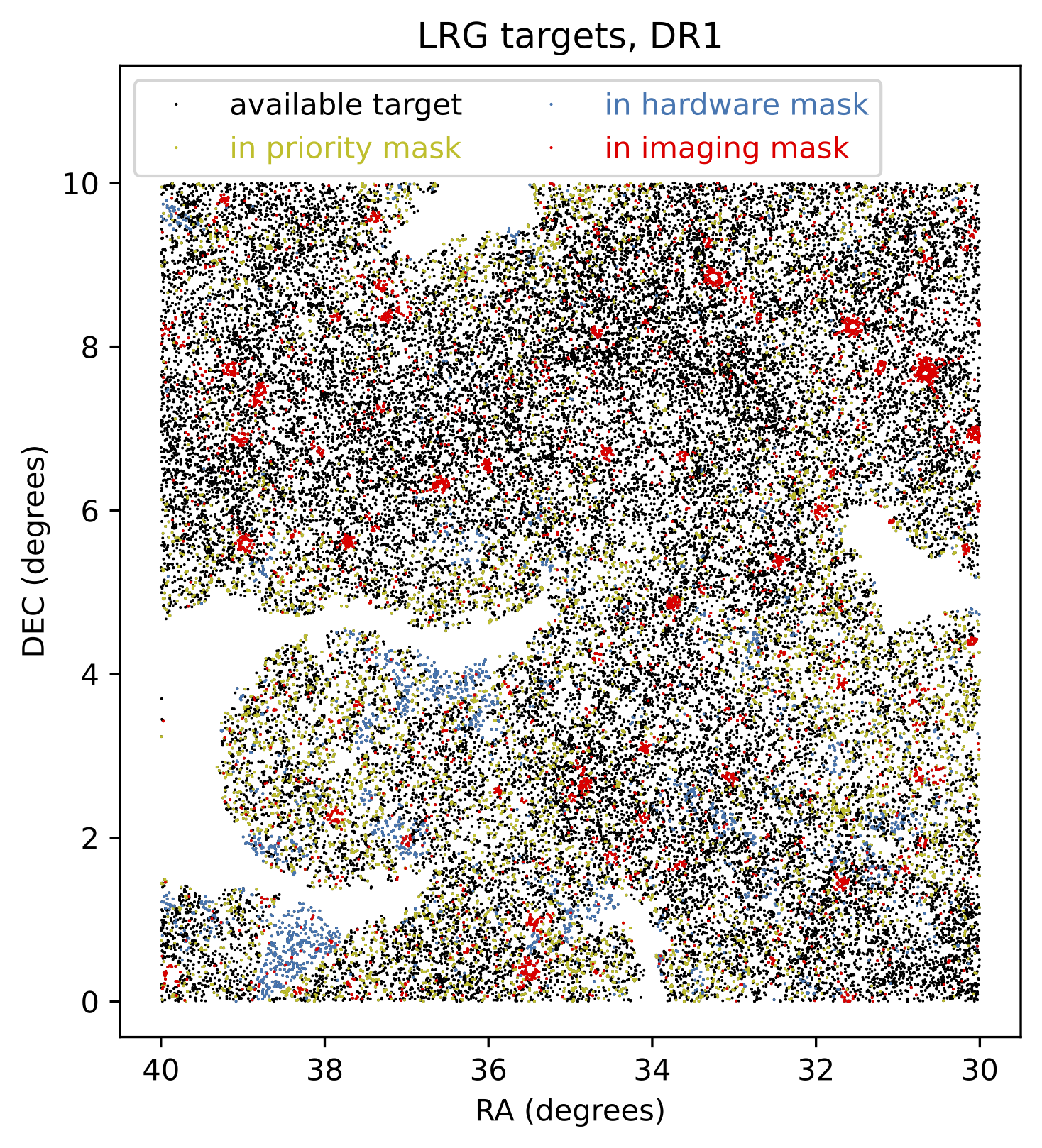}
    \includegraphics[width=0.47\columnwidth]{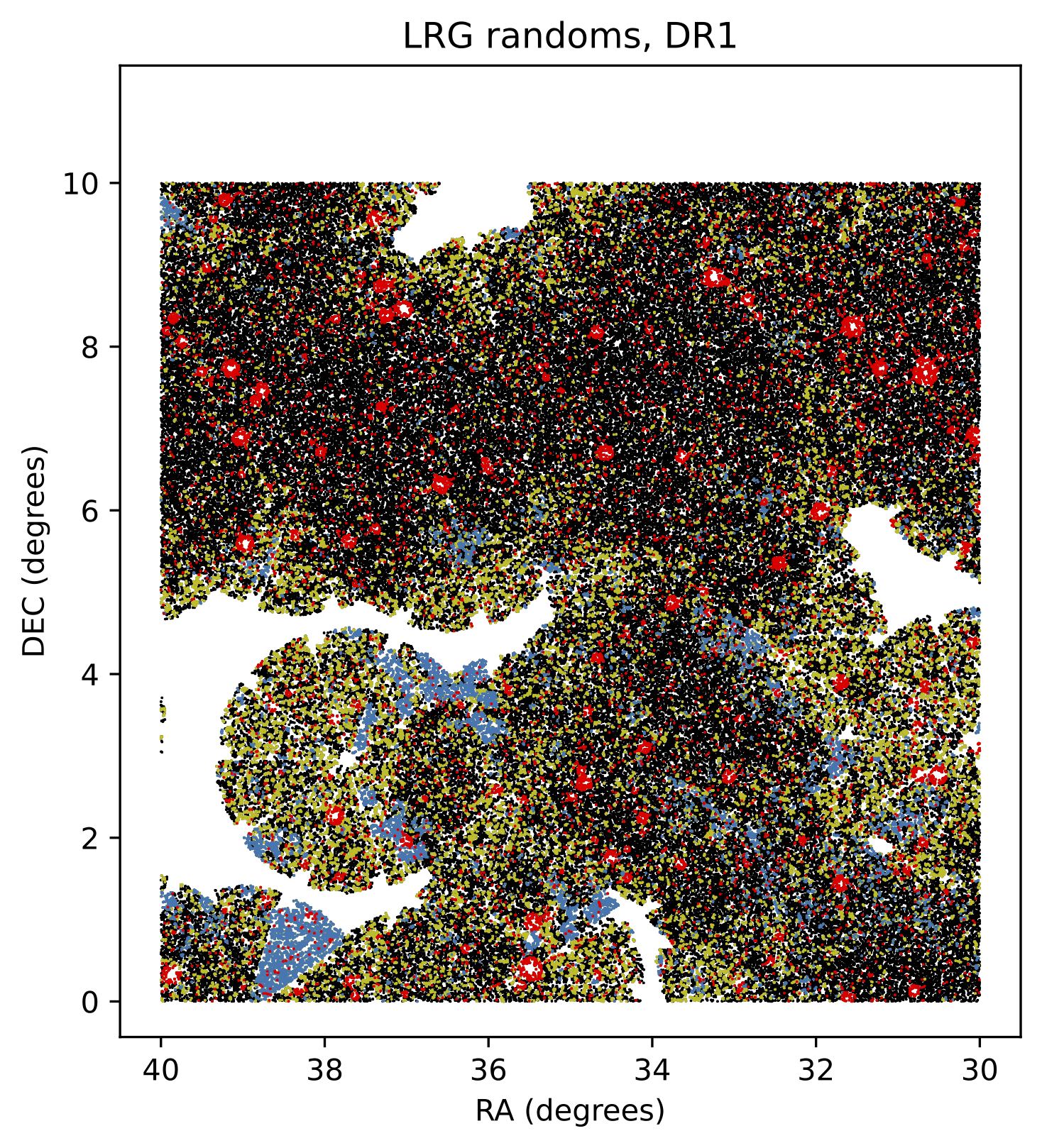}
    \includegraphics[width=0.47\columnwidth]{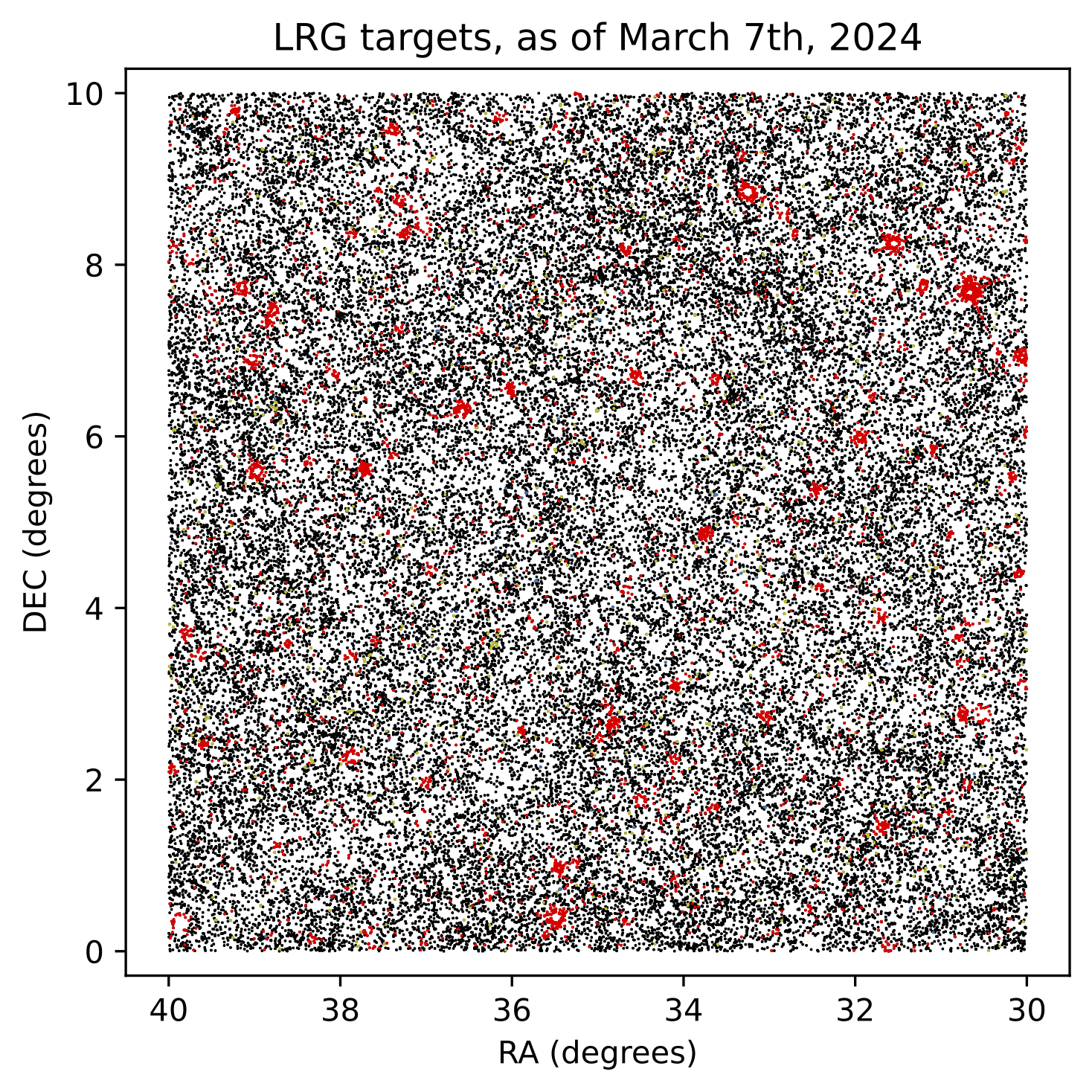}
    \includegraphics[width=0.47\columnwidth]{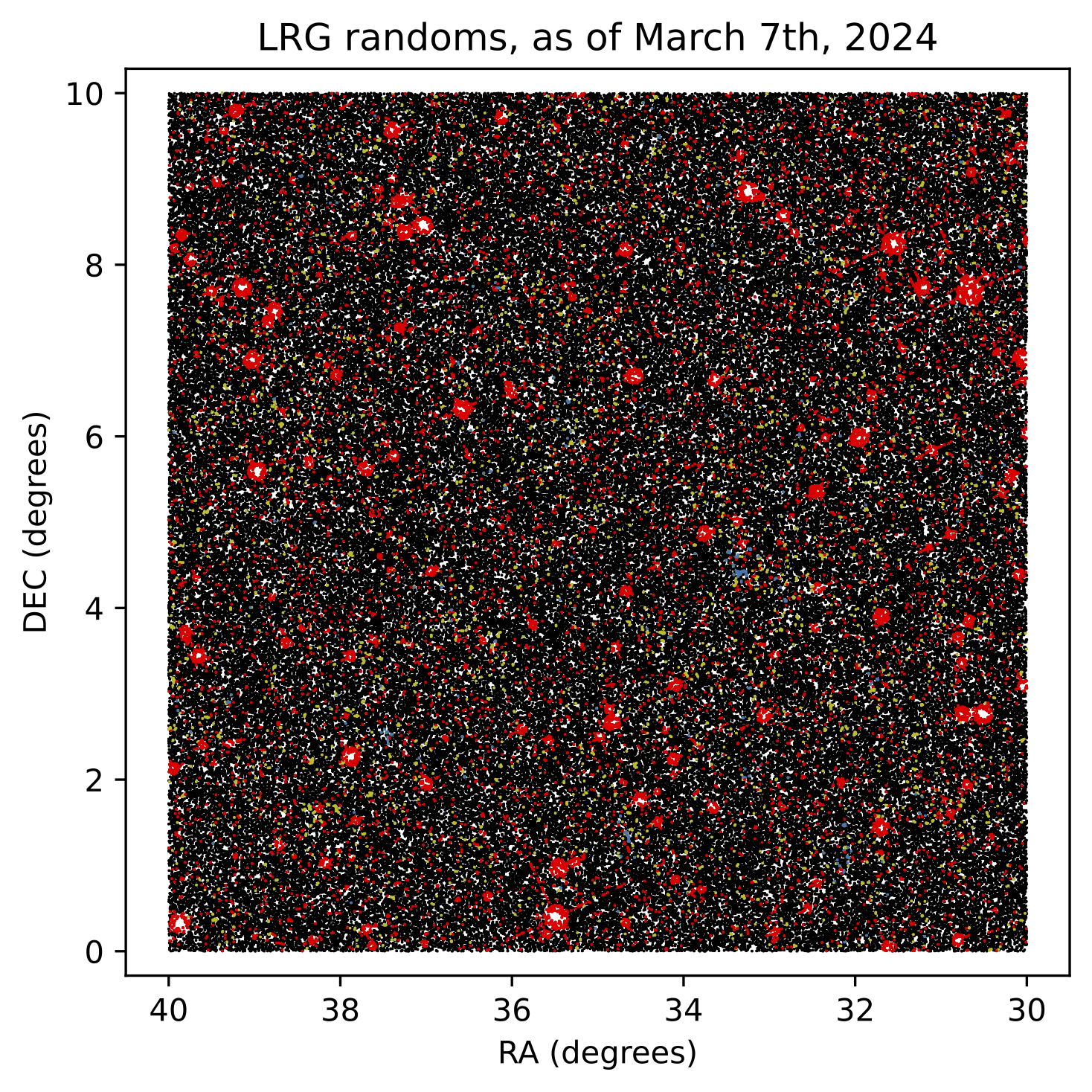}
   \caption{Available LRG targets (left) and randoms (right) in a small area of the DESI footprint. The top panels show the results in the DR1 catalogs and the bottom panels show the results as of March 7th, 2024. Various veto masks are indicated. The unmasked data is shown in black. One can observe matching patterns in data and randoms, as intended.}
    \label{fig:mask}
\end{figure*} 

Fig. \ref{fig:mask} shows the angular effect of veto masks applied to the DR1 LRG data and random samples, within a small area of the footprint. Unmasked data can be seen with black points. The yellow points are areas vetoed due to the priority of what was assigned blocking the possibility of any LRG assignments. This primarily happens in areas that have only been covered by one tile; one can observe that significant amounts of such areas are yellow. The blue points denote data where some aspect of the DESI instrument was flagged as poorly performing and thus denoted `bad hardware' (i.e., $H_{\rm gd} = \texttt{False}$, as defined above). In the lower left-hand corner of each of the upper panels, one can observe that an entire petal was so flagged. Looking again at Fig. \ref{fig:NTILE}, one can see no data in the location for DR1 and slightly lower coverage than average as of March 20th, 2024. This is because data with $H_{\rm gd} = \texttt{False}$ does not contribute to our coverage determination. At other locations, smaller-sized patterns are removed; these can correspond to, e.g., amplifier boundaries on the CCDs that read out the information from the spectrograph or groups of 50 fibers that failed to move properly due to canbus communication errors. Similar to the priority mask, such areas exist primarily where only one tile has covered the area. For the vast majority of the combination of exposures and fibers the DESI hardware performs well and failure is thus especially unlikely for any area covered more than once. In total, in DR1, only 3\% and 2\% footprint is vetoed due to poorly performing hardware, for dark- and bright-time respectively, and these numbers are expected to decrease in future data releases as the coverage increases (when the same area is covered again, it is unlikely the hardware will again be poorly performing, especially given it will be a different area in the focal plane).
 All of this information is so tracked and incorporated into the LSS catalog footprint.

In the DR1 analysis, we make additional cuts that remove the area in the extrema of imaging property maps, whose creation we describe in Section \ref{sec:imagingsys}. The goal is to remove only a small amount of the footprint but ensure that non-representative data do not impact the regressions. The choices applied to DR1 are given in \cite{DESI2024.II.KP3} and in total remove 3.4\% of the footprint. We expect the particular cuts and the fractional area removed to be updated with each data release. We add `\_HPmapcut' to the file name for these catalogs. Thus, as of DR1, for `full' catalogs we have:

\textbullet~ `full\_noveto'; the catalogs that contain everything reachable, as determined by \textsc{fiberassign}

\textbullet~ `full'; the catalogs that have good hardware, priority, and imaging conditions vetos applied

\textbullet~ `full\_HPmapcut'; the catalogs that have an additional cut applied based on imaging properties.

\subsection{Assignment Completeness Definitions}
\label{sec:comp}
 As of DR1, we determine all assignment completeness statistics using the `full' catalogs, i.e., those determined before applying the image property map cuts. Statistics are determined at the target-sample level, i.e., a complete sample would be one where all of the targets of the type under consideration had been assigned a fiber, without any consideration of any results obtained from the measured spectra. Thus, the total completeness, $C_{\rm tot}$ that we determine, for any set of targets, is the number of those targets assigned to a fiber divided by the total number within the set. 

 As of DR1, we have divided the fiber assignment incompleteness into two components, in a way at mimics the SDSS approach. They are:

\textbullet~ $f_{\rm TLID}$ (column named \texttt{FRACZ\_TILELOCID}) This is computed per \texttt{TARGETID} and is the inverse of the number of targets sharing the same \texttt{TILELOCID}, $N_{\rm TLID}$, (and thus all of the targets with the same \texttt{TILELOCID} will have the same value). Only one of the targets at a given \texttt{TILELOCID} could have been observed. The $N_{\rm TLID}$ value is essentially the number of targets that were competing for the fiber. Thus, $f_{\rm TLID}$, can be thought of as a per \texttt{TILELOCID} completeness, and its inverse can be used as a completeness weight. Such a weight is analogous to the close-pair weights used in SDSS. It accounts for all fiber assignment incompleteness due to competition for fibers between targets of the same type. As of DR1, the values are determined before applying vetos. The good hardware and priority vetos are associated with \texttt{TILELOCID} and the thus order makes no difference for those vetos and $f_{\rm TLID}$ values. However, applying the imaging veto after determining $f_{\rm TLID}$ implies that targets inside of the imaging veto mask with a \texttt{TILELOCID} matching that of an assigned target outside of the veto mask will count in the statistics. We will revisit this choice in future data releases.

\textbullet~ $f_{\rm tile}$ (column named \texttt{FRAC\_TLOBS\_TILES}) 
This second component of the assignment completeness accounts for any incompleteness not accounted for by $f_{\rm TLID}$. More specifically, it accounts for the fraction of potential assignments that were not associated with fibers that were observed by the target type and were not rejected by the priority veto. This could be due to, e.g., a standard star or sky fiber needing to be assigned to meet the minimum threshold.  Such assignments are not deterministic; all targets are randomly assigned a \texttt{SUBPRIORITY}, that provides a relative preference for assignment for targets that have the same priority values. Standards and skies are thus most likely to take an assignment from the targets that have (randomly) been assigned the lowest \texttt{SUBPRIORITY}. Competition with other target types can also contribute, especially for ELGs, which compete with LRGs (and most LRGs have a higher priority for assignment). However, the ELGs that are promoted to LRG priority are done so at random. Thus, in all of these cases, the process should be random (in terms of the sky area affected), within the given set of overlapping tiles and such targets would not be accounted for with $f_{\rm TLID}$. All of the unassigned targets should have a non-zero probability of assignment, provided a different random seed when producing DESI target catalogs. 

We  determine $f_{\rm tile}$ by calculating the fraction of targets that are at a \texttt{TILELOCID} assigned to the given target type (e.g., LRG), determined for each unique set of overlapping tiles, $t_{\rm group}$ (recorded as \texttt{TILES} in the catalogs; see Section \ref{sec:count}). The size of any individual $t_{\rm group}$ will depend on the geometry of the overlapping tiles. Any change of the color in Fig. \ref{fig:NTILE} represents a change in the $t_{\rm group}$ (as the number of over-lapping tiles is constant within a $t_{\rm group}$, by definition).  
Recall that in Section \ref{sec:full_noveto}, we defined the boolean quantity $T_{\rm fa}$ that is \texttt{True} if the given \texttt{TILELOCID} was assigned to the given target type (but not necessarily the particular target). 
Thus, for any selection of data, this completeness is $\sum T_{\rm fa}/N_{\rm tot}$. Phrased differently, it is $(N_{\rm assigned}+N_{\rm associated})/N_{\rm tot}$, where $N_{\rm associated}$ is the number of targets not assigned but at \texttt{TILELOCID} that were assigned to the given target type (and $N_{\rm tot}$ is the total number of targets of the given type). We determine it for each unique $t_{\rm group}$ and it is thus
\begin{equation}
f_{\rm tile}(t_{\rm group}) = \frac{\sum_{i\in t_{\mathrm{group}}}{T_{{\mathrm{fa},i}}}}{N_{\rm tot}(t_{\rm group})}.
\label{eq:ftile}
\end{equation}
Importantly, the randoms contain the same $t_{\rm group}$ information as the data (which were used to determine the number of overlapping tiles mapped in Fig. \ref{fig:NTILE}) and we thus add $f_{\rm tile}(t_{\rm group})$ to the randoms after determining the values for the data. Each unique group of overlapping tiles ($t_{\rm group}$) is analogous to an SDSS `sector' and $f_{\rm tile}$ is thus analogous to $C_{\rm BOSS}$.

The total assignment completeness is simply the number of assigned targets divided by the total number of targets. Per unique $t_{\rm group}$ we also record the total assignment completeness, $C_{\rm tile}$ (column named \texttt{COMP\_TILE}). This quantity is used to present maps of the DR1 completeness in \cite{DESI2024.II.KP3}.

The combination of $f_{\rm tile}$ and $f_{\rm TLID}$ provide the total assignment completeness. We thus expect that if one selects data that has been assigned ($L_{\rm fa} = \texttt{True}$) the sum of their inverse should be the total number of targets in the full catalog, i.e., we expect $\sum 1/f_{\rm tile}[L_{\rm fa}=\texttt{True}]1/f_{\rm TLID}[L_{\rm fa}=\texttt{True}] = N_{\rm tot}$. 

\section{Modelling Selection Function Variations}
\label{sec:weights}
As of DR1, we have identified three primary effects that we must model in order to properly determine the variations in the DESI selection function (within the observed area defined in the previous sections). They are all accounted for in the catalogs by applying weights to either data or randoms. The process thus makes the expected ratio of data to random (after weighting) constant\footnote{For any effect that is possible to account for with weights on the randoms, one could equivalently sub-sample the randoms rather than weight them.}. These are variations in completeness, target sample properties due to imaging properties, and the ability to measure good redshifts due to the effective exposure time of DESI observations or properties concerning DESI hardware (e.g., the location on the focal plane). We will go through each in this section, paying particular attention to the technical details of how they can be accounted for in the DESI LSS catalog structure. Particular analysis choices for the DR1 LSS catalogs are detailed in \cite{DESI2024.II.KP3} and at most only briefly described here. Weights to optimize the signal-to-noise of clustering observations will be discussed in Section \ref{sec:clustering}.

\subsection{Assignment Completeness}
\label{sec:comp_weight}
For assignment completeness weights, the approach is to correct for the probability of assignment. By construction, the correct target density can be recovered by taking the observed number of targets and dividing it by the assignment completeness. The same holds for any sub-selection of the data (e.g., by redshift), on average, except the noise increases. Thus, any one-point statistic can be determined without bias by weighting by an accurate inverse probability of assignment. One could thus use $C_{\rm tot} = f_{\rm TLID} f_{\rm tile}$ and $w_{\rm comp} = 1/C_{\rm tot}$. For many quick calculations, this is indeed used. However, applying $f_{\rm tile}$ as a weight on the randoms provides the same density contrast. This has better noise properties in most cases and, e.g., it is applied to the randoms in the DR1 LSS `clustering' catalogs, described in Section \ref{sec:clustering}. We expect the $p_{\rm obs}$ probability determined from the alternative MTL realizations (described below) to be a statistical match to $f_{\rm TLID} f_{\rm tile}$. The completeness weights are discussed further in Section \ref{sec:clustering} and are always taken from the $f_{\rm TLID}, f_{\rm tile}, p_{\rm obs}$ quantities available in the full catalogs.

Properly accounting for the assignment incompleteness effects on clustering measurement requires simulations of the assignment histories. To break ties in the assignment of fibers to targets of the same class, every DESI target has a \texttt{SUBPRIORITY} that is a random number between 0 and 1, and the target with the greatest value is assigned the fiber. Alternate realizations of the DESI survey assignment history can thus be obtained by randomizing the \texttt{SUBPRIORITY} values and re-running the fiber assignment. To do so properly, assignments must follow the same process as applied to the `merged target ledger' (MTL), which tracks the assignment history of every DESI target, updating information after every observation. The process of simulating DESI fiber assignment after randomizing the \texttt{SUBPRIORITY} values is thus referred to as the `alternative MTL' process (`altmtl'), as each realization represents a separate MTL history. This process as developed for DESI SV3 and DR1 LSS catalogs is described fully in \cite{KP3s7-Lasker}. Many realizations can be used to record whether or not a given target was assigned in each realization with a bit map, which can then be used to measure unbiased 2-point functions via pair-wise inverse probabilities \cite{2017Bianchi}. Importantly, in such a case, $f_{\rm tile}$ should not be applied to the randoms, and the random redshifts (see Section \ref{sec:randomz}) should be sampled accounting for the individual inverse probability weights, $1/p_{\rm obs}$, for each observed target, given by the altmtl.

\subsection{Density Variations due to Imaging}
\label{sec:imagingsys}
The DESI target samples can have variations in their number density caused by changes in the conditions of the imaging used to define their selection, both in the total projected target density and in the shape of the ${\rm d}N/{\rm d}z$. If untreated, these variations will cause spurious clustering results to be obtained from the DESI LSS catalogs. The variations thus must be modelled and corrected with some treatment applied to either the data or randoms.

A now standard approach is to perform regressions between the observed density and maps of imaging properties. To employ this approach, the values of image properties are first attached to the DESI `full' random catalogs. This is done at the finest resolution available for the given map (e.g., the imaging depth is sampled from the sub-arcsecond-sized pixels that make up the LegacySurvey imaging data). We then average the values put into the random files within \textsc{Healpix} \citep{healpix} pixels Nside=256 (resolution value). The process of producing these maps closely follows that for the `{\it pixelized weight maps}' described in section 4.5.2 of \cite{DESItarget}. The maps are produced separately for the data in the Northern (BASS/MzLS) and Southern (DECam) imaging regions and are used to perform regressions to obtain weights for systematic variation in target densities. Further details on the maps used for DR1 analysis are provided in \cite{DESI2024.II.KP3}.

Given the imaging property maps, many regression methods exist for modelling trends with the observed DESI density. As of DR1, three have been incorporated into the DESI LSS pipeline. For each, the final result is a weight attached to the observed redshift that nulls the observed trends. These are:
\begin{itemize}
    \item The neural net method \textsc{sysnet}\footnote{
https://github.com/mehdirezaie/sysnetdev}, most recently applied to \cite{Rezaie23} and denoted \texttt{WEIGHT\_SN} in the catalogs;
\item The random forest method \textsc{regressis}\footnote{https://github.com/echaussidon/regressis}, first applied in \cite{Chaussidon21QSOsys}  and denoted \texttt{WEIGHT\_RF} in the catalogs;
\item The linear method applied to eBOSS LSS catalogs \cite{BautistaDR14LRG,ebosslss} (\texttt{WEIGHT\_IMLIN} in the catalogs).
\end{itemize}
For each, the density contrast between data and randoms in the `full' catalogs is determined and is then regressed against the selected imaging property masks. For the data, good spectroscopic observations in particular redshift ranges are selected and completeness weights are applied, e.g., for DR1 LRGs, we perform the regressions separately in the three redshift bins used in the clustering analysis: $0.4<z<0.6$, $0.6<z<0.8$, $0.8<z<1.1$. The weights are then included in the `full' data catalogs based on the true redshift. This approach allows any redshift-dependent corrections to be applied before the catalog-level blinding (as applied in DR1) shifts the redshifts \cite{KP3s9-Andrade}. The full details on their implementation and results for DR1 are presented in \cite{DESI2024.II.KP3}.  

The regression methods described will only correct the projected density (within whatever chosen redshift bin) and will not account for any variations that are smooth with redshift and thus cause variation in the normalized ${\rm d}N/{\rm d}z$. This can be included as an additional weight that depends on redshift as well as the map features. The impact of fitting for and including such a weight for the DR1 ELG sample is studied in \cite{KP3s2-Rosado}.

Ideally, none of the weights described here would be necessary and we could instead forward-model the target sample selections and thus include all fluctuations in the randoms. The forward-modelling technique \textsc{Obiwan} \citep{Kong20} has indeed shown promise in simulating DESI ELG \cite{KP3s2-Rosado} and LRG \cite{KP3s13-Kong} samples. This has made it a vital tool in explaining the observed variation and identifying what must be due to calibration issues (especially of the Galactic extinction corrections). It is not yet at the level where the ${\rm d}N/{\rm d}z$ at a given sky location can be predicted at the accuracy (in comparison to the data) and precision (limited by the number of simulations run) required for use as randoms. If sufficient improvements in the forward modelling are achieved, the results could be used either as weights to apply to the data or directly to the randoms, to correct trends with imaging properties.

\subsection{Redshift Failures}
Similar to the weights that account for trends between the observed DESI density and the imaging properties of the target samples, we perform regressions on the data to uncover trends between the DESI observing properties and the redshift success rate. The model parameters of the regressions are determined by selecting DESI data of the given tracer type observed with good hardware (i.e., $H_{\rm gd}={\rm True}$; see Section \ref{sec:vetos}) and features relating to the DESI observation. The model can then be applied to the full catalogs to add new columns for the predicted redshift success and the weight, $w_{\rm zf}$ (\texttt{WEIGHT\_ZFAIL} in the catalogs), to account for relative fluctuation in the predicted success. 

We expect the dominant source of variation in redshift success rate, $f_{\rm zgood}$ to be the `template signal to noise ratio squared', $S_{\rm ratio}$ (\texttt{TSNR2\_(tracer)} in the catalogs). This is determined per observation, using a constant template spectrum; it is thus equivalent to an effective observing time. See \cite{DESIpipe} for details. We further expect $f_{\rm zgood}$ to be monotonic with $S_{\rm ratio}$. The redshift success rate for DESI tracers as a function of effective observing time in early DESI observations can be found in each of the target selection studies for each respective tracer BGS, QSO, LRG, ELG:  \cite{BGStarget,QSOtarget,LRGtarget,ELGtarget}.

The $f_{\rm zgood}$ trends with $S_{\rm ratio}$ are further expected to be greatest for the targets with the least fiberflux. For ELGs, the overlap between the observed \otwo wavelength and sky lines as a function of redshift introduces a strong redshift dependence that must be taken into account. Thus, for DR1, the weights for BGS, LRG, and QSO are a function of $S_{\rm ratio}$ and fiberflux, while for ELGs, they are a function of $S_{\rm ratio}$ and redshift. The overall process of accounting for these redshift success trends in the DR1 data is described in \cite{KP3s3-Krolewski} and the specific concerns of ELGs (including any residual dependency on fiberflux) are addressed in \cite{KP3s4-Yu}.

Regression against further metadata, such as the spectrograph number, the focal plane location, and the given positioner's accuracy could be added to provide a more complete model for the expected redshift success rate. For DR1, trends with these metadata (and more) are investigated and their impacts on clustering measurements are demonstrated in \cite{KP3s3-Krolewski} and \cite{KP3s4-Yu} and found to be negligible. However, the significance of the underlying trends in the data suggests the results should be re-evaluated with every data release.

As of DR1, we are only able to account for trends with redshift failures by applying weights to the data. To instead apply the results to randoms, each random point would need to be correctly associated with a particular observation, per tracer. In the DR1 randoms, we only determine whether the point could have been assigned to the given tracer type, and when choosing the particular associated observation we choose the one with the greatest $S_{\rm ratio}$. It is likely possible to improve this by considering the `pass' (the $N$th time the area has been covered, one through up to seven) and the probability of assigning a given \texttt{TILELOCID} to a given target type (e.g., using the altmtl results).

\section{Enabling Null Tests and Density Determination}
\label{sec:null}
After the weights for the selection function have been determined and added to the `full' catalogs, numerous null tests can be performed and the number density as a function of redshift, $n(z)$, can be calculated. Typical null tests include measuring the density of a DESI sample as a function of some observational property (e.g., effective observing time) or foreground map (e.g. Galactic extinction) and verifying that the variation is consistent with the statistical noise. Passing such tests validates the selection function modeling. Providing this ability on the full catalogs is important for DESI analysis, as it allows the catalogs to be validated and $n(z)$ determined independently of the `clustering' catalogs that are directly used for clustering measurements (and discussed in the following section). This allows catalog-level blinding, such as done for DR1 \cite{KP3s9-Andrade}, to be performed that does not interfere with the validation process. All of the DR1 work on the selection function and its validation (\cite{DESI2024.II.KP3,KP3s3-Krolewski,KP3s4-Yu,KP3s2-Rosado}) was performed on the unaltered data. However, any effect on the clustering measurements was demonstrated using the blinded catalogs, up until the analysis was frozen and the DESI collaboration unblinded the results.

Providing the $n(z)$ for the unaltered data is important, e.g., for ensuring that simulations reproduce properties of the measured data. In DR1, the blinding method shifted redshifts coherently, to alter any inferred distance scales \cite{KP3s9-Andrade}. Therefore, the blinded catalogs could not be used (directly) for $n(z)$ measurements. The $n(z)$ can be calculated for any selected region by counting galaxies weighted by their assignment completeness, thereby providing the expected $n(z)$ for a sample with no assignment incompleteness. The volume the galaxies occupy can be determined by assuming a cosmological model and calculating the volume of the spherical shell given the redshift bounds. This can then be multiplied by the fractional sky coverage. The area of the sky can be determined by counting the number of randoms in the LSS catalog and dividing by the density of randoms, which is 2500 deg$^{-2}$ per DESI LSS random file.

\section{Clustering Catalogs}
\label{sec:clustering}
 The DESI LSS `clustering' catalogs are cut to good spectroscopic observations and the redshift range intended for clustering analysis. They only include the columns required to calculate clustering statistics. 

\subsection{Reducing Data to what is used for Clustering}
\label{sec:goodz}
For both data and randoms, the clustering files take the `full' LSS catalogs and reduce them to a subset of columns necessary for calculating clustering statistics\footnote{Any later desired columns can be obtained via a \texttt{TARGETID} match to the `full' catalogs.}. The full LSS catalogs contain no cuts based on the outcome of the redshift pipeline. We cut to only those objects with good redshifts to produce the clustering catalogs. They will further be cut to the redshift range of interest and/or where we are confident we have properly modelled the selection function. In this subsection, we describe the process applied to the data.

 The LSS catalog construction code can trivially accommodate any choice based on the properties of the redshift fit. Two quantities likely to always be relevant for selecting good redshifts are the redshift pipeline flag \texttt{ZWARN} and the $\Delta\chi^2$~(\texttt{DELTACHI2}) obtained from the redshift fitting pipeline between the best and next-best-fit redshifts. For DR1, good redshifts are decided based on the same criteria as for the SV3 catalogs described in \cite{EDR} and these choices match the recommendations of the respective targeting catalogs \citep{LRGtarget,ELGtarget,QSOtarget,BGStarget}. The data are also cut to the redshift range within which we have modelled the selection function. The details of the selection applied to DR1 can be found in \cite{DESI2024.II.KP3}.

For the data, each type of weight to account for the selection function, described in the subsections of Section \ref{sec:weights}, is condensed to a single weight, $w_{\rm comp}$ (\texttt{WEIGHT\_COMP}) for completeness, $w_{\rm sys}$ (\texttt{WEIGHT\_SYS}) for target density fluctuations due to imaging conditions, and $w_{\rm zfail}$ (\texttt{WEIGHT\_ZFAIL}) to account for changes in the relative redshift success rates. The weights are combined to produce the total weight, $w_{\rm tot}$ (\texttt{WEIGHT}), i.e.:
\begin{equation}
w_{\rm tot} = w_{\rm comp}w_{\rm sys}w_{\rm zfail}.
\end{equation}
The $w_{\rm tot}$ will later be re-scaled (coherently for data and randoms) based on the completeness variations, as described in Section \ref{sec:FKP}. While the above description currently only applies to DR1, we expect it to remain true throughout future DESI releases. If, e.g., we account for variations with imaging systematics by imparting variations into the randoms, we would simply set $w_{\rm sys}=1$ for all of the data.

All weights are kept in the `clustering' catalogs to make it easy to test the impact of ignoring a given weight (e.g., one would divide the $w_{\rm tot}$ column by $w_{\rm zfail}$ when running clustering measurements to these the effect of ignoring redshift failure weights). Similarly, when multiple versions of the regressions against imaging properties have been run, we provide each of the individual weights for testing. Additionally, we keep the number of overlapping tiles (\texttt{NTILE}) to allow tests of clustering as a function of coverage. The comprehensive list with definitions will be updated with each data release.\footnote{
\url{https://desidatamodel.readthedocs.io/en/latest/DESI\_ROOT/vac/RELEASE/lss/VERSION/LSScats/clustering/clustering\_dat.html}} Any desired columns not provided can be obtained via a \texttt{TARGETID} match to the `full' catalogs (or any other DESI catalogs).

\subsection{Adding Radial Information to Randoms}
\label{sec:randomz}
As of DR1, we assign radial information to the random catalogs by sampling points from the data. This will provide a perfect match between the observed ${\rm d}N/{\rm d}z$ of the data and the randoms. This perfect match comes at a cost, as purely radial modes of clustering will be nulled and the effect must be accounted for in any analysis. We refer to this as the radial integral constraint \cite{demattiaIC}. This is a practical choice based on the fact that true ${\rm d}N/{\rm d}z$ cannot be known from first principles or even easily approximated by simple functions of redshift.\footnote{Though, see \cite{RossMGS} for an exception.} If the ${\rm d}N/{\rm d}z$ of DESI data can be shown to be accurately predicted in the future (e.g., from first principles and forward modeling), the choice of sampling from the data can be easily revisited within the LSS pipeline.

Radial information is added to the randoms separately in each of the relevant photometric regions, based on how the target selection was split. For BGS, ELG, and LRG samples, the target selection was defined separately in the DECam (`S') and BASS/MzLS (`N') regions. For QSO, the `S' region was further divided into DES and notDES regions. In DR1, we simply randomly sample the data within each respective region when assigning redshifts to the randoms in each region. We sample the redshifts, associated weights, and any other properties (e.g., photometry) that we might later want to use to produce subsamples from the data to assign them to the randoms. The \texttt{TARGETID} of the data is also sampled (and stored as \texttt{TARGETID\_DATA}), to keep a record and also enable further columns to be added later if desired. Such sampling ensures that the weighted ${\rm d}N/{\rm d}z$ of the randoms matches that of the data in each region, and will continue to match as long as one cuts both the data and randoms on the same properties. If the definition of $w_{\rm comp}$ used for the data is only $1/f_{\rm TLID}$, as in DR1, the random $w_{\rm tot}$ then gets multiplied by $f_{\rm tile}$ (to account for incompleteness that was not yet accounted for via weights on the data).

We then normalize such that the weighted ratio of data to randoms is the same in each photometric region. We do so by multiplying the $w_{\rm tot}$ of the randoms by the required factor. The data and random that were split by region are concatenated into one catalog. At this point, both the data and random clustering catalogs have four fundamental columns that can, e.g., be passed to clustering estimators: \texttt{RA}, \texttt{DEC}, \texttt{Z}, \texttt{WEIGHT}, which are, respectively, the angular celestial coordinates, the redshift, and the combined weight, $w_{\rm tot}$.  In DR1, we then split them again by Galactic hemisphere, `NGC' and `SGC'. The following subsection describes how we re-factor $w_{\rm tot}$ and add an additional weight for consideration, to optimize the signal-to-noise of clustering measurements.

\subsection{Weighting to Optimally Balance Sampling Rate}
\label{sec:FKP}

Given a local density and a power spectrum amplitude, \cite{FKP} describe how to optimally weight galaxies for power spectra measurements. It is common practice to assume some fiducial power spectrum amplitude, $P_0$, and then only apply a weight, $w_{\rm FKP}$ that is based on how the number density changes with redshift. The DESI DR1 sample has large completeness variations that are primarily driven by the number of overlapping tiles, $N_{\rm tile}$. 
Therefore, we developed a method that uses an estimate of the density as a function of both redshift and the number of overlapping tiles, $n(z,N_{\rm tile})$, which can be used in the standard `FKP' weight formula:
\begin{equation}
w_{\rm FKP} = 1/(1+n(z,N_{\rm tile})P_0).
\end{equation}

To use the above definition for $w_{\rm FKP}$, we first factor the completeness from the data and randoms, to account for the fact that we are already up-weighting data based on completeness. We do so by considering the completeness and related weights as a function of the number of overlapping tiles, $N_{\rm tile}$. We calculate $\langle w_{\rm comp}\rangle (N_{\rm tile})$, $\langle f_{\rm tile}\rangle (N_{\rm tile})$, and $\langle C_{\rm tile}\rangle (N_{\rm tile})$. To account for the fact that galaxies have been up-weighted by $w_{\rm comp}$, we divide $w_{\rm tot}$ for both data and random by $\langle w_{\rm comp}\rangle (N_{\rm tile})$, i.e.,
\begin{equation}
w_{\rm tot}(N_{\rm tile}) = w_{\rm tot}/\langle w_{\rm comp}\rangle (N_{\rm tile}).
\end{equation}
The same quantity $\langle w_{\rm comp}\rangle (N_{\rm tile})$ is used to determine $n(z,N_{\rm tile})$, given the already estimated $n(z)$ for a complete sample (defined in Section \ref{sec:null}):
\begin{equation}
n(z,N_{\rm tile}) = n(z)\langle f_{\rm tile}\rangle (N_{\rm tile})/\langle w_{\rm comp}\rangle (N_{\rm tile}).
\end{equation}
The $n(z,N_{\rm tile})$ values for each data and random point are stored as the column \texttt{NX} in the catalogs. In general, we expect that the highest signal-to-noise clustering measurements will be obtained by weighting both data and randoms by $w_{\rm tot}w_{\rm FKP}$. However, this could depend on the specific type of clustering measurement or the scales being used and one could potentially obtain a better weight for their analysis that replaces $w_{\rm FKP}$ (for both data and randoms). The $w_{\rm tot}$ should always be used as is for data and randoms, with possible exceptions described in the following subsection.

\subsection{Producing Subsamples}
\label{sec:subsamples}

One can select samples from the data and randoms, applying the same selections to both, using any of the columns in the clustering catalogs other than the weights. A caveat is that the $w_{\rm sys}$ may no longer be appropriate (and using them as is may indeed bias the clustering results) if one cuts on photometric properties. In such a case, one could determine new $w^{\prime}_{\rm sys}$, and then multiply the existing $w_{\rm tot}$ for both data and random\footnote{The $w^{\prime}_{\rm sys}$ can be attached to the randoms based on the \texttt{TARGETID\_DATA} attached when sampling data redshifts } by $w^{\prime}_{\rm sys}/w_{\rm sys}$. 

The BGS sample, in particular, is one where results will be obtained by splitting it in many different ways. Further, given its selection is flux-limited, the sample characteristics will evolve strongly with redshift. In DR1, a choice was made to simplify the analysis of the BGS sample by applying an $r$-band absolute magnitude threshold, based on $k$-corrections determined by \textsc{FastSpecFit} \cite{fastspecfit,fastspecfit_code} and an $e$-correction matching that applied to the EDR \cite{EDR}. Full details are provided in \cite{DESI2024.II.KP3}. For this sample, we chose to apply the cut on the `full' catalogs and then proceed through the same LSS catalog pipeline steps as any other sample (i.e., starting with the selection function determination and including the assignment of radial information to the randoms). One advantage of proceeding in this fashion was that we preserved the sky density of the randoms; i.e., no photometric cut needed to be applied to the randoms and their density was able to remain 2500 deg$^{-2}$. A similar process can be applied to any sample selected from the `full' catalogs, assuming any geometrical component can be equally applied to the randoms.

\section{Summary}
\label{sec:con}
We have described the technical framework for producing DESI LSS catalogs. The linchpin of the process is the ability to use the DESI \textsc{fiberassign} software to track the sky geometry where DESI observations were possible to an accuracy of better than one arcsecond. Further, with every observed DESI tile, we evaluate the hardware performance at a level as fine as individual fibers. This combination allows us to account for the irregular and dynamic shape of the footprint associated with any observed DESI tile. See Fig. \ref{fig:tilegeo}. 

\begin{figure*}
    \centering 
    \includegraphics[width=0.95\columnwidth]{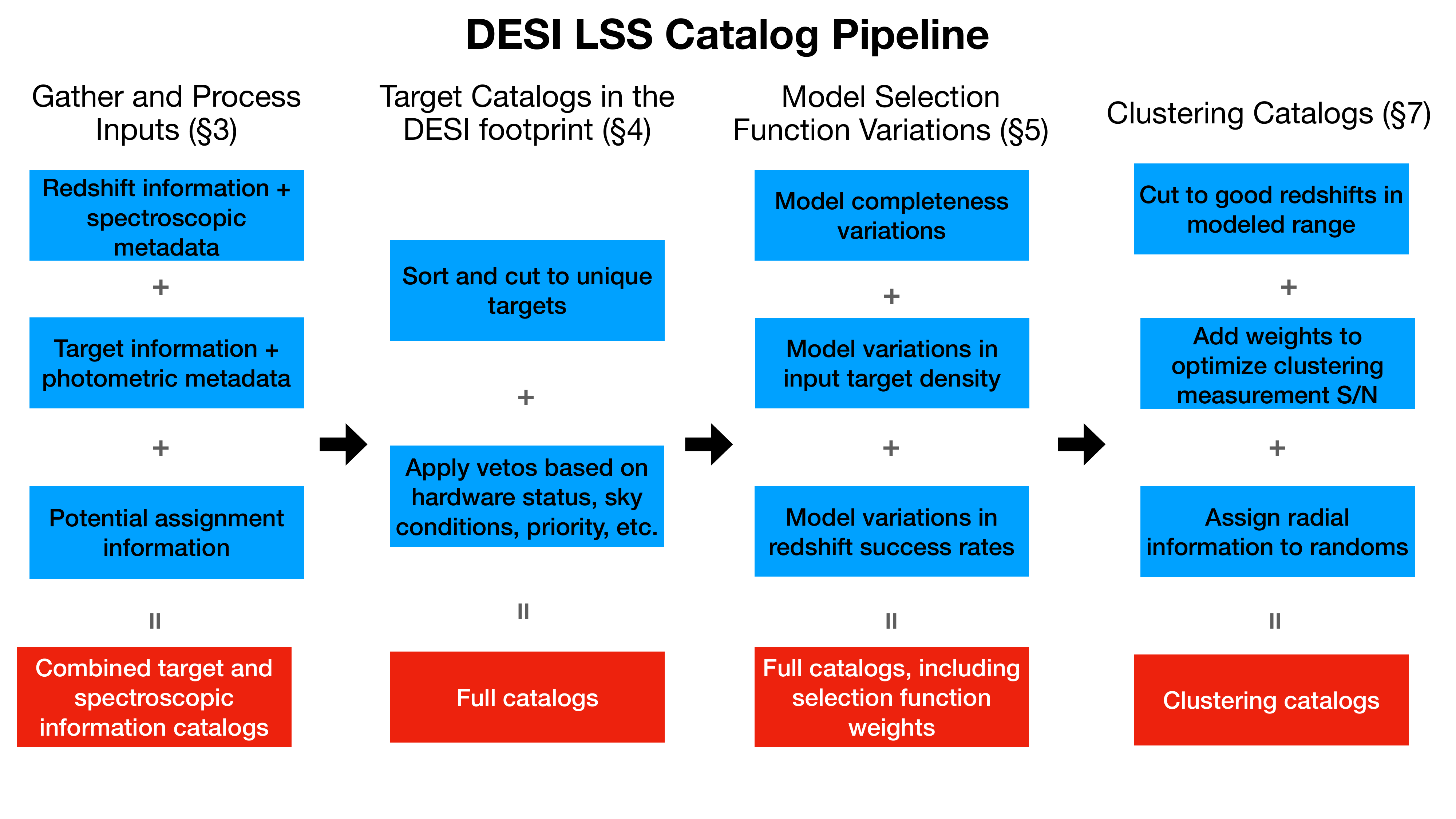}
   \caption{A flowchart detailing the DESI LSS catalog pipeline construction steps. The full context, including key definitions, can be found within the labeled sections of this paper.}
    \label{fig:flow}
\end{figure*}

A flowchart detailing the steps in the DESI LSS catalog pipeline is shown in Fig. \ref{fig:flow}. We summarize these steps as:
\begin{itemize}
    \item {\bf 1) Collect all information on reachable targets and randoms.} This step is detailed in Section \ref{sec:inputs}. The information associated with every instance where a target or random was reachable is combined. Each unique \texttt{ TARGETID} can appear up to the number of overlapping tiles at its sky location. Information from the associated spectroscopic observations is matched to this data, as is all information relevant to veto masks. After removing data associated with poorly performing hardware, quantities such as the number of overlapping tiles can be determined. See Fig. \ref{fig:NTILE}.
    \item {\bf 2) Reduce targets and randoms to unique entries.} This step is detailed in Section \ref{sec:full} and produces what we refer to as the `full' catalogs. The results from step 1) are carefully sorted to keep the most relevant \texttt{ TARGETID} occurrence when cutting to unique entries, as described in Section \ref{sec:full_noveto}. The result is unique entries for data and randoms, for every target that was reachable to a DESI fiber positioner (whether or not it was assigned and without regard for the results of the redshift fitting pipeline), including all veto mask information. The area within different types of veto masks is shown in Fig. \ref{fig:mask}. Veto masks can then be applied, as described in Section \ref{sec:vetos}, and then the completeness statistics defined in Section \ref{sec:comp} can be determined.
    \item {\bf 3) Model selection function variations.} The observed density of DESI data will vary within the footprint due to at least three known effects: fiber assignment incompleteness, fluctuations in the target sample inherited from properties of the imaging data, and fluctuations in the DESI redshift success rate due to properties of DESI observations and/or hardware components. In Section \ref{sec:weights}, we detail how these effects can be modelled and validated using the `full' catalogs, typically with the use of weights. We further describe where the current framework allows weights to be applied to data and/or randoms and the basics of the approach applied to DR1 (with full details on the DR1 approach contained in \cite{DESI2024.II.KP3}). Section \ref{sec:null} then discusses how validation tests can be applied to the `full' catalogs.
    \item {\bf 4) Produce catalogs intended for clustering measurements.} Given modelling of the selection function, clustering statistics accounting for all identified effects can be calculated. To simplify this, we produce `clustering' catalogs that include only the information relevant to producing the clustering measurement, as detailed in Section \ref{sec:clustering}. Any catalog-level blinding that is to be applied happens between the full and clustering catalog stages. First, the data clustering catalogs are created by cutting to `good' redshifts within the range of interest. The selection function weights are combined into a total \texttt{WEIGHT} column. Radial information is then added to the random catalogs, matching the weighted ${\rm d}N/{\rm d}z$ of the data (described in Section \ref{sec:randomz}). We then apply weights to both the data and randoms that are meant to maximize the expected signal to noise, given the changes in the observed number density as a function redshift and number of overlapping tiles (described in Section \ref{sec:FKP}). Finally, we provide information on how one could further divide the clustering catalogs.
\end{itemize}

This work describes the status of the DESI LSS catalog pipeline, as of DR1. Particular analysis choices for DR1 are detailed in \cite{DESI2024.II.KP3}. We expect that there will be many improvements applied to future DESI data releases, but these will mainly pertain to the particular analysis choices. We expect that the framework we have developed and the steps outlined above will remain nearly constant throughout the lifetime of DESI. However, some changes, corrections, and/or additions to the pipeline are inevitable and these will be detailed in the corresponding future data releases and/or LSS catalog papers.  

\section{Data Availability}
The plotting scripts and data used to generate all of the plots in this paper will be uploaded to \url{https://doi.org/10.5281/zenodo.11193828}. 

\acknowledgments

This material is based upon work supported by the U.S. Department of Energy (DOE), Office of Science, Office of High-Energy Physics, under Contract No. DE–AC02–05CH11231, and by the National Energy Research Scientific Computing Center, a DOE Office of Science User Facility under the same contract. Additional support for DESI was provided by the U.S. National Science Foundation (NSF), Division of Astronomical Sciences under Contract No. AST-0950945 to the NSF’s National Optical-Infrared Astronomy Research Laboratory; the Science and Technology Facilities Council of the United Kingdom; the Gordon and Betty Moore Foundation; the Heising-Simons Foundation; the French Alternative Energies and Atomic Energy Commission (CEA); the National Council of Humanities, Science and Technology of Mexico (CONAHCYT); the Ministry of Science and Innovation of Spain (MICINN), and by the DESI Member Institutions: \url{https://www.desi.lbl.gov/collaborating-institutions}. Any opinions, findings, and conclusions or recommendations expressed in this material are those of the author(s) and do not necessarily reflect the views of the U. S. National Science Foundation, the U. S. Department of Energy, or any of the listed funding agencies.

The DESI Legacy Imaging Surveys consist of three individual and complementary projects: the Dark Energy Camera Legacy Survey (DECaLS), the Beijing-Arizona Sky Survey (BASS), and the Mayall z-band Legacy Survey (MzLS). DECaLS, BASS and MzLS together include data obtained, respectively, at the Blanco telescope, Cerro Tololo Inter-American Observatory, NSF’s NOIRLab; the Bok telescope, Steward Observatory, University of Arizona; and the Mayall telescope, Kitt Peak National Observatory, NOIRLab. NOIRLab is operated by the Association of Universities for Research in Astronomy (AURA) under a cooperative agreement with the National Science Foundation. Pipeline processing and analyses of the data were supported by NOIRLab and the Lawrence Berkeley National Laboratory. Legacy Surveys also uses data products from the Near-Earth Object Wide-field Infrared Survey Explorer (NEOWISE), a project of the Jet Propulsion Laboratory/California Institute of Technology, funded by the National Aeronautics and Space Administration. Legacy Surveys was supported by: the Director, Office of Science, Office of High Energy Physics of the U.S. Department of Energy; the National Energy Research Scientific Computing Center, a DOE Office of Science User Facility; the U.S. National Science Foundation, Division of Astronomical Sciences; the National Astronomical Observatories of China, the Chinese Academy of Sciences and the Chinese National Natural Science Foundation. LBNL is managed by the Regents of the University of California under contract to the U.S. Department of Energy. The complete acknowledgments can be found at \url{https://www.legacysurvey.org/}.

Any opinions, findings, and conclusions or recommendations expressed in this material are those of the author(s) and do not necessarily reflect the views of the U. S. National Science Foundation, the U. S. Department of Energy, or any of the listed funding agencies.

The authors are honored to be permitted to conduct scientific research on Iolkam Du’ag (Kitt Peak), a mountain with particular significance to the Tohono O’odham Nation.

\appendix

\section{Glossary}

A glossary of DESI quantities and jargon is available at \url{https://data.desi.lbl.gov/doc/glossary/}. Below, we repeat some of the entries that are most relevant for the LSS catalogs and provide entries for terms and quantities defined by the LSS process.

\noindent\textbullet ~{\bf BGS} `Bright Galaxy Sample'; Galaxies targeted during bright time (see below); the sample is primarily flux-limited \cite{BGStarget}.

\noindent\textbullet~ {\bf bright time}: DESI observations taken during `bright' conditions, as defined in \cite{SurveyOps.Schlafly.2023}. BGS are the only LSS targets during bright time.

\noindent\textbullet~ {\bf dark time}: DESI observations taken during `dark' conditions, as defined in \cite{SurveyOps.Schlafly.2023}. ELG, LRG, QSO are observed during dark time.

\noindent\textbullet~ \textsc{\bf \textsc{desitarget}}: The code package used to select targets for DESI spectroscopic observation and change their status based on their observation history; \cite{DESItarget} \url{https://github.com/desihub/desitarget}.

\noindent\textbullet ~{\bf ELG} `Emission Line Galaxy'; a class of DESI targets (see below) selected with the expectation they will yield a detection of OII flux with redshift $z$ between 0.6 and 1.6. The selection is defined in \cite{ELGtarget}.

\noindent\textbullet~ \textsc{\bf fiber}: An individual fiber optic from a positioner on the focal plane to a spectrograph. Fibers are numbered sequentially from 0 to 4999, corresponding to value of \texttt{FIBER} in the catalogs.

\noindent\textbullet~ \textsc{\bf fiber positioner}: A two arm moveable robot holding a DESI fiber on the focal plane.

\noindent\textbullet~ \textsc{\bf fiberassign}: The code package used to determine which targets can be assigned to which fiber on the DESI focal plane; \cite{FBA.Raichoor.2024}, \url{https://github.com/desihub/fiberassign}.

\noindent\textbullet~ {\bf Legacy Surveys}: The program that delivered the photometric information used to select targets for DESI spectroscopy, via their data release 9 (DR9); \cite{LS.Overview.Dey.2019,LS.dr9.Schegel.2024}, \url{https://www.legacysurvey.org/dr9/}.

\noindent\textbullet~ {\bf \texttt{LOCATION}}: The identifier corresponding to a particular fiber positioner. Each \texttt{LOCATION} value has a one-to-one mapping to each \texttt{FIBER} value.

\noindent\textbullet ~{\bf LRG}: `Luminous Red Galaxy'; a class of DESI targets distinguished by their red colours, resulting from a strong 4000 angstrom break.

\noindent\textbullet~ {\bf MTL}: The `Merged Target Ledger' contains all of the information on how the state of a target has changed. It is updated through \textsc{desitarget} (see above) and controls its priority in \textsc{fiberassign} (see above). Primarily, a target will go from unobserved (and thus high \texttt{PRIORITY}; see below) to observed (and thus low \texttt{PRIORITY}). 

\noindent\textbullet ~{\bf \texttt{PRIORITY}}: The quantity given to targets to determine the relative preference for assigning a fiber. The initial \texttt{PRIORITY} are determined based on the target type and the values are reduced after a successful observation.

\noindent\textbullet ~{\bf QSO}: Technically `Quasi-Stellar Object', but we use it synonymously with `quasar'; a class of DESI targets likely to be quasars \cite{QSOtarget}. Those with redshifts $>$ 2.1 are 'Lyman-$\alpha$' quasars, which are at high enough redshift to allow measurement of 'Lyman-$\alpha$' forest absorption. 

\noindent\textbullet~ {\bf random}: Object with celestial coordinates randomly selected at a uniform density from a specified region on the sky.

\noindent\textbullet~ {\bf target}: Object selected via photometry for DESI spectroscopic followup by \textsc{desitarget} (see above), but not necessarily observed (yet) by DESI. Each has a unique \texttt{TARGETID}. Similarly, randoms (see above) that only occupy sky locations where there was Legacy Survey (see above) DR9 imaging were produced by \textsc{desitarget} and have unique \texttt{TARGETID}.

\noindent\textbullet~ {\bf tile}: A single DESI pointing on the sky with assignments of which fibers should observe which targets; each has a unique \texttt{TILEID}.

\noindent\textbullet~ \texttt{TILES} A string listing the tiles that the target appeared on, using the \texttt{TILEID}s sorted in ascending order and separated by `-'. Each unique \texttt{TILES} represents a unique group of overlapping tiles.

\noindent\textbullet~ {\bf \texttt{TILELOCID}}: The identifier we use to match to information associated with a particular tile and fiber, defined as 10000\texttt{TILEID}+\texttt{LOCATION}.

\noindent\textbullet~ {\bf \texttt{TSNR2}}: Template Signal-to-Noise Squared. A signal-to-noise metric weighted by what wavelengths matter most for determining the redshift of DESI targets, given their magnitude and redshift distributions. This depends upon target class, e.g. Lyman-alpha QSO TSNR2 more heavily weights blue wavelengths, while ELG TSNR2 more heavily weights redder wavelengths which cover the emission lines for the DESI redshifts of interest. TSNR2 depends upon the noise properties of individual spectra, but not the signal properties of the target. It is fully defined in \cite{DESIpipe}.

\bibliographystyle{JHEP}
\bibliography{references,DESI2024_updated30Apr}

\end{document}